\documentclass[fleqn,usenatbib]{mnras}

\usepackage{newtxtext,newtxmath}

\usepackage[T1]{fontenc}

\DeclareRobustCommand{\VAN}[3]{#2}
\let\VANthebibliography\thebibliography
\def\thebibliography{\DeclareRobustCommand{\VAN}[3]{##3}\VANthebibliography}

\usepackage{graphicx}	
\usepackage{amsmath}	
\usepackage{caption}

\title[CO foreground removal by UNet]{Foreground Removal of CO Intensity Mapping Using Deep Learning}

\author[Zhou, X. et al. ]{
Xingchen Zhou,$^{1,2}$
Yan Gong,$^{1,3}$\thanks{E-mail: gongyan@bao.ac.cn}
Furen Deng,$^{1,2}$
Meng Zhang,$^{1,2}$
Bin Yue,$^{1}$ 
and Xuelei Chen$^{1,2,4,5}$
\\
$^{1}$National Astronomical Observatories, Chinese Academy of Sciences, 20A Datun Road, Beijing 100101, China\\
$^{2}$University of Chinese Academy of Science, Beijing, 100049, China\\
$^{3}$Science Center for China Space Station Telescope, National Astronomical Observatories, Chinese Academy of Sciences,\\ 20A Datun Road, Beijing 100101, China \\
$^{4}$Department of Physics, College of Sciences, Northeastern University, Shenyang 110819, China \\
$^{5}$Center for High Energy Physics, Peking University, Beijing 100871, China
}

\date{Accepted XXX. Received YYY; in original form ZZZ}

\pubyear{2015}

\begin{document}
\label{firstpage}
\pagerange{\pageref{firstpage}--\pageref{lastpage}}
\maketitle

\begin{abstract}
	Line intensity mapping (LIM) is a promising probe to study star formation, the large-scale structure of the Universe, and the epoch of reionization (EoR). Since carbon monoxide (CO) is the second most abundant molecule in the Universe except for molecular hydrogen ${\rm H}_2$, it is suitable as a tracer for LIM surveys. However, just like other LIM surveys, CO intensity mapping also suffers strong foreground contamination that needs to be eliminated for extracting valuable astrophysical and cosmological information. In this work, we take $^{12}$CO($\it J$=1-0) emission line as an example to investigate whether deep learning method can effectively recover the signal by removing the foregrounds. The CO(1-0) intensity maps are generated by N-body simulations considering CO luminosity and halo mass relation, and we discuss two cases with median and low CO signals by comparing different relations. We add foregrounds generated from real observations, including thermal dust, spinning dust, free-free, synchrotron emission and CMB anisotropy. The beam with sidelobe effect is also considered. Our deep learning model is built upon ResUNet, which combines image generation algorithm UNet with the state-of-the-art architecture of deep learning, ResNet. The principal component analysis (PCA) method is employed to preprocess data before feeding it to the ResUNet. We find that, in the case of low instrumental noise, our UNet can efficiently reconstruct the CO signal map with correct line power spectrum by removing the foregrounds and recovering PCA signal loss and beam effects. Our method also can be applied to other intensity mappings like neutral hydrogen 21cm surveys.
\end{abstract}




\section{Introduction}
\label{sec:introduction}
Line intensity mapping (LIM) is a promising technique that can efficiently explore huge survey volume in an affordable observational time.
It integrates the emission line intensity from galaxies and intergalactic medium (IGM), and does
not require resolving individual sources~\citep{Visbal2011,Gong2011,Kovetz2017,Fonseca2017,Bernal2022}.
LIM has been widely proposed in studying the epoch of reionization (EoR), galaxy star formation, cosmic large-scale structure, etc. It offers an opportunity to comprehensively explore our Universe in redshifts beyond the reach of traditional galaxy surveys.

Several emission lines are commonly proposed and used as tracers in LIM, such as carbon monoxide (CO), singly ionized carbon [CII], H$\alpha$, and neutral hydrogen 21cm emission lines~\citep{Visbal2010,Gong2011,Carilli2011,Lidz2011,Gong2012,Gong2013,Silva2013,Pullen2014,Uzgil2014,Gong2014,Silva2015,Fonseca2017,Gong2017,Gong2020}.
Among these lines, CO is the second most abundant molecule in the Universe, except for molecular hydrogen (H$_2$).
At high redshift ($z\gtrsim6$) around EoR, CO can trace the star-forming regions within galaxies to probe cosmic reionization history \citep{Gong2011,Carilli2011,Lidz2011}. At low redshift with $z\lesssim3$, CO LIM can be used to investigate star formation rate, and map the large-scale matter field to study the nature of dark energy and dark matter \citep{Kovetz2017}.
The CO rotational line emission is at ladder of frequency $\nu_{J\rightarrow J-1} = J\times{\rm 115.27GHz}$
for $J\rightarrow J-1$ transition. Since CO lines are usually bright in galaxies, ground-based telescopes targeting sub-mm wavelength can efficiently observe them~\citep{Bernal2022}.
Currently, several ongoing and planning probes are targeting at CO lines, such as COPSS~\citep{Keating2015,Keating2016}, mmIME~\citep{Keating2020}, COMAP~\citep{Cleary2022,Breysse2022}, FYST~\citep{Aravena2021} and SPT-SLIM~\citep{Karkare2022}.
COMAP recently has published their first-stage results and shown promising prospects~\citep{Ihle2022,Chung2022}.

On the other hand, foreground removal is a big problem in CO and other LIM surveys, that can greatly affect the signal extraction. The LIM foregrounds include the continuum components, such as thermal dust, free-free and synchrotron emissions, and non-continuum
ones, e.g. CMB anisotropy, spinning dust emission, and contaminations of other emission lines from different redshifts at the same observed frequency.
The continuum foreground usually is attempted to be removed by blind subtraction methods, such as principal component
analysis (PCA)~\citep{Abdi2010,Wold1987}, fast independent component analysis (fastICA)~\citep{Hyvarinen2000}, etc.
PCA is one of the mostly used methods. It can extract the principal components with large eigenvalues of covariance
matrix, and reduce the contamination by removing the contribution from these modes. Although PCA is effective in removing continuum foregrounds, it may corrupt the information and cause signal loss at large scales~\citep{Alonso2014,Deng2022}. So algorithms that can recover the power spectrum usually need to be applied after PCA subtraction.

The instrumental effects will further complicate the foreground removal of LIM. Gaussian white noise induced by system temperature of telescope usually can affect the signal extraction from observational maps. Although it is relatively easy to be evaluated and removed in power spectra since it should be a constant at different scales, this noise should be suppressed as much as possible to avoid overwhelming the signal. Another instrumental effect called beam effect should also be carefully considered.
The beam effect is caused by main and side lobes of radio antenna~\citep{Matshawule2021}, and has frequency-dependent feature. On one hand, it can affect the spatial resolution and result in a drop in the power spectrum at small scales. On the other hand, beam effect can change the shape of foreground spectrum in frequency space. Hence, the foreground removal by blind subtraction methods, e.g. PCA, usually cannot obtain satisfactory performance when beam effect is considered~\citep{Ni2022}.

Deep learning is widely used in information extraction of images, that probably can be applied in foreground removal and instrumental effect correction in LIM. Deep learning is one of the most prosperous branches of machine learning. The development and optimization of deep learning algorithm are in active research.
Convolutional neural networks (CNN) introduced by~\citet{Fukushima1982} are the most widely used architecture.
The performance of CNN improves with increase of convolutional layers, however, deep layers will suffer gradient vanishing problem and
cannot learn effectively. ~\citet{He2015} proposes ResNet that can address this problem using identity mapping. Applying ResNet block, CNN can be built very deep, and even hundreds of convolutional layers will not cause gradient vanishing problem, greatly improving the learning ability of CNN.

For image generation tasks like foreground removal, UNet and generative adversarial network (GAN) are the suitable networks.
UNet is orginated from autoencoder~\citep{Rumelhart1985} and improve its generative ability through concatenation with features obtained in encoding path~\citep{Ronneberger2015}. Autoencoders aim to transform inputs to outputs with small distortions. However, with skip connection introduced in UNet and deep convolutional layers, the transformation ability of autoencoder significantly increases.
This structure can be used for segmentation of filaments, walls and voids in large scale structure~\citep{Aragon2019}, mapping galaxies from dark matter halos~\citep{Zhang2019,Kasmanoff2020}, removal of foreground~\citep{Makinen2021,Ni2022} and many other interesting applications in astronomical and cosmological field.

GAN is a generative model proposed in recent years \citep{Goodfellow2014}, and it is more complicated than common networks.
It has a critic or discriminator that tells the realness of generated images, and guides the generator to output more
realistically.
One structure of GAN family that can perform image-to-image generation is called Pix2PixGAN~\citep{Isola2016}. This GAN can be regarded as conditional GAN (cGAN)~\citep{Mirza2014} with input images as conditions.
GAN structures can simulate realistic cosmic web in extremely short time~\citep{Rodriguez2018}, generate 3D N-body
simulations~\citep{Perraudin2019,Feder2020}, and polish simulations from existing low resolution ones~\citep{Kodi2020,Li2021}.
Therefore it can be built into emulators, and hopefully substitute the expensive N-body or hydrodynamical
simulations. Denoising LIM is also a promising application~\citep{Moriwaki2021,Moriwaki20212}.

In this work, we take $^{12}$CO($\it J$=1-0) line as an example, which has a rest-frame wavelength of 2.6 mm and frequency of $\rm 115.27\ GHz$, to study foreground removal using deep learning.  We consider low-redshift CO(1-0) line intensity mapping at $z\sim1$ contaminated by foregrounds and instrumental noise, including thermal dust, spinning dust, free-free emission, synchrotron emission, CMB anisotropy, beam effect, thermal noise, etc. CO line intensity maps are mocked by N-body simulation and the relation of CO luminosity and halo mass. Foreground components at frequencies of CO signals are obtained from PySM3 python library, which are fitted by publicly available data from WMAP and Planck surveys and extended to small scale~\citep{Thorne2017}.
The instrumental effects we consider are based on COMAP, and the Bessel beam model is assumed in the analysis. PCA method is tested and used as preprocessing for input of our UNet, and finally we evaluate the foreground removal performance by UNet.

The paper is organized as follows: we discuss our mock data including CO intensity maps, foregrounds, instrumental noise and beam effect in Section~\ref{sec:data}. In Section~\ref{sec:methods}, we introduce our deep learning model and training process.
The results are shown in Section~\ref{sec:result} and we also have some discussions on instrumental and PCA effects. We conclude this work in Section~\ref{sec:conclusion}.

\section{Mock Data}
\label{sec:data}

We consider CO(1-0) intensity mapping at $z\sim1$ contanminated by several foreground components, and instrumental effects including white noise and beam effects induced by radio telescope are also simulated. We use a N-body simulation at redshift around $z=1$ to generate our CO mock maps. The foreground components considered are thermal dust, spinning dust, synchrotron, free-free emissions and CMB anisotropy. The white noise is simulated based on and modified from instrumental parameters of COMAP, and the Bessel beam model is employed to mock the beam effects of radio antenna.

\subsection{CO intensity maps}
\label{sec:CO maps}

We generate CO mock maps based on MDPL2 from CosmoSim simulation sets~\citep{Klypin2016}. This N-body simulation uses cosmological parameters from Planck2013~\citep{Ade2014} and is performed by L-Gadget2 code~\citep{Springel2005}. The box size is 1 ${\rm Gpc}/{\rm h}$, number of particles is $3840^3$, with a mass resolution of $1.51\times10^9\, h^{-1}{\rm M_{\odot}}$ of single dark matter particle. This resolution is large enough for our LIM study.
We make use of the halo catalogue obtained by friend-of-friend (FOF) method from snapshot of $z=1.03$ to be our basis for CO intensity maps. We split the whole box to subboxes of 300 ${\rm Mpc}/{\rm h}$ to generate lightcones with angles of $7.45\times7.45 \deg^2$. After generating the lightcones, we slice the planes according to the resolution of frequency, taken to be 0.1 GHz in this work. Since we set $z=1$ at the middle of subbox, the frequency range is from 55.2 to 60.2 GHz with 0.1 GHz spacing.
We find $N$ halos inside each spatial pixel in each frequency plane, and calculate the observed CO intensity by:
\begin{equation}
	\label{eq:observed CO intensity}
	I_{\rm CO}(\nu) = \frac{1}{(\Delta\theta)^2}\sum_i^N\frac{1}{\Delta\nu}\frac{L^i_{\rm CO}}{4\pi r^2_i(1+z_i)^2},
\end{equation}
where $\Delta\theta$ is the angular size of spatial pixel, $\Delta\nu$ is the resolution of frequency, $L_{\rm CO}^i$, $r_i$ and $z_i$ are CO luminosity, comoving distance and redshift of the $i$th halo, respectively. The CO luminosity can be calculated from $L_{\rm CO}-M_{\rm h}$ relation, and we obtain this relation with the help of infrared (IR) luminosity.

\begin{figure}
	\includegraphics[width=0.95\columnwidth]{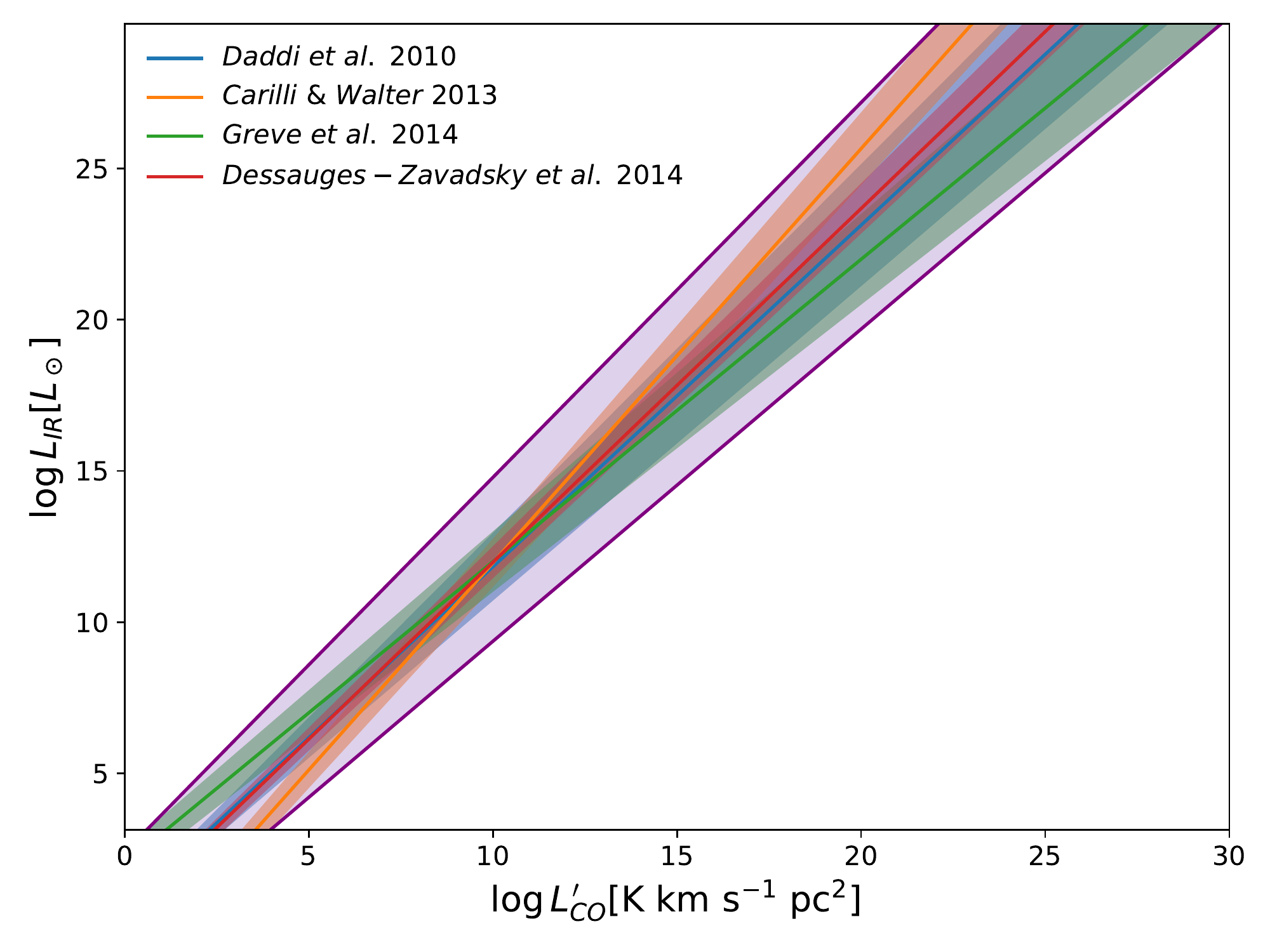}
	\caption{$L_{\rm IR} - L_{\rm CO}^\prime$ relations used in this work. We consider the red line with $(\alpha,\beta)=(1.17,0.28)$ as our fiducial model \citep{Dessauges2015}. The lower and upper bounds of these luminosity relations (purple lines) are also derived with $(\alpha,\beta)=(1.24, 2.39)$ and $(1.03, -0.96)$, respectively.}
	\label{fig:IR_CO_relation}
\end{figure}

Since CO and IR luminosity are both star formation tracers, they can be closely related by empirical relation:
\begin{equation}
	\label{eq:CO and IR relation}
	\log(L_{\rm IR}) = \alpha\log(L_{\rm CO}^\prime) + \beta,
\end{equation}
where $L_{\rm IR}$ is in units of $L_\odot$ and $L_{\rm CO}^\prime$ is in units of ${\rm K\ km\ s^{-1}\ pc^2}$.
$\alpha$ and $\beta$ are fitted from galaxy observations, and we consider $(\alpha,\beta)=(1.13\pm0.09, 0.53\pm0.22)$ \citep{Daddi2010}, $(1.37\pm0.04, -1.74\pm0.40)$ \citep{Carilli2013}, $(1.00\pm0.05, 2.00\pm0.50)$ \citep{Greve2014}, and $(1.17\pm0.03, 0.28\pm0.23)$ \citep{Dessauges2015} in this work. These relations are measured by observations, and are available at $L_{\rm IR}\lesssim 10^{14}\ \rm L_{\sun}$ \citep[see e.g.][]{Aravena2012}, which should cover the halo mass range with $M\lesssim10^{15}\ \rm M_{\sun}$ in our simulation. The relations are shown in Figure~\ref{fig:IR_CO_relation}, and the lower and upper bounds of the  luminosity relations are also estimated as $(\alpha,\beta)=(1.24, 2.39)$ and $(1.03, -0.96)$, respectively, shown in purple lines. We take the median values $(\alpha,\beta)=(1.17, 0.28)$ as our fiducial model and the lower bound with $(1.24, 2.39)$ as the low CO signal model to check the performance of our foreground removal method.
The unit conversion from $L_{\rm CO}^\prime$ to $L_{\rm CO}$ in units of $L_\odot$ is
\begin{equation}
	\label{eq:unit conversion}
	L_{\rm CO} = 4.9\times10^{-5}L_\odot\left(\frac{\nu_{\rm CO}^r}{115.27{\rm GHz}}\right)^3
	\left(\frac{L_{\rm CO}^\prime}{{\rm K\ km\ s^{-1}\ pc^2}}\right),
\end{equation}
where $\nu_{\rm CO}^r = 115.27{\rm GHz}$ is the rest-frame frequencies of CO(1-0) transition.

\begin{figure}
	\includegraphics[width=0.95\columnwidth]{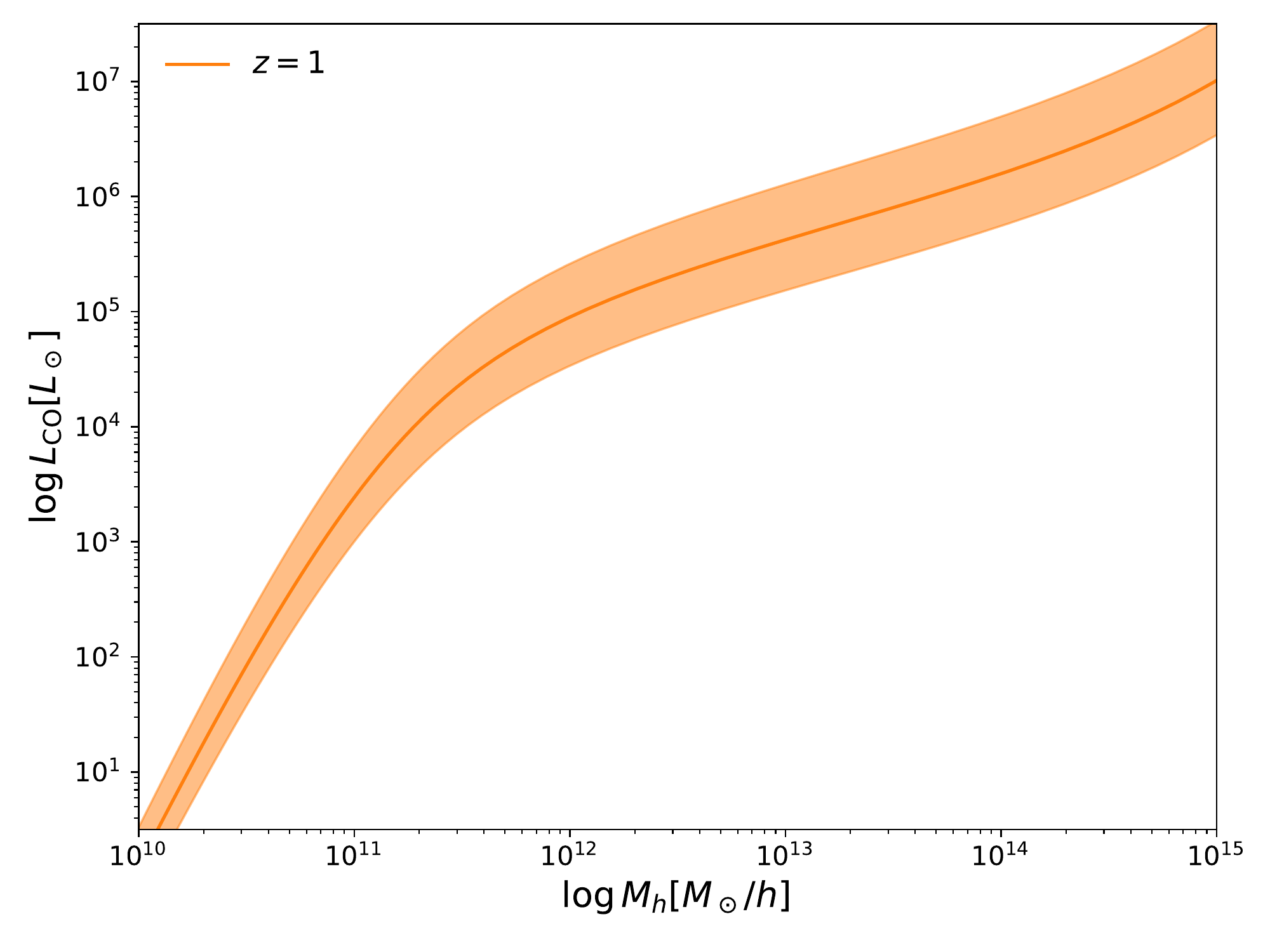}
	\caption{$L_{\rm CO}-M_{\rm h}$ relation at $z=1$ derived from our fiducial $L_{\rm IR} - L_{\rm CO}^\prime$ model. The scatter of the relation is also shown in orange bands, which is estimated by the uncertainties of $\alpha$ and $\beta$ in the CO and IR luminosity relation.}
	\label{fig:CO_Hm relation}
\end{figure}

Then we need to derive the relation between IR luminosity $L_{\rm IR}$ and halo mass $M_{\rm h}$. By assuming the "abundance matching" technique can be applied, this relation could be obtained. Here we also assume that the subhalos contribute little to the IR luminosity and can be neglected~\citep{Yue2015}. The abundance matching equates the number density of galaxies with IR luminosity above a luminosity $L_{\rm IR}$ and the number density of halos above a halo mass $M_{\rm h}$ by
\begin{equation}
	\label{eq:abundance matching}
	\int_{> L_{\rm IR}}\Phi(L_{\rm IR}, z){\rm d} L_{\rm IR} = \int_{>M_{\rm h}}\frac{{\rm d}n}{{\rm d}M_{\rm h}}{\rm d}M_{\rm h},
\end{equation}
where $\Phi(L_{\rm IR}, z)$ is the IR luminosity function, and ${\rm d}n/{\rm d}M_{\rm h}$ is the halo mass function given in \citet{Sheth1999}. Based on the simulation, we set the ranges of the halo mass and IR luminosity as $M_{\rm}\in(10^{10}, 10^{15}\, {\rm M}_{\sun}/h)$ and $L_{\rm IR}\in(10^5, 10^{14}\,{\rm L}_{\sun})$, respectively, to perform the abundance matching.

The IR luminosity functions of spiral galaxies, starburst galaxies, and star-forming galaxies including AGNs are comprehensively studied in literatures \citep[e.g.][]{Magnelli2009,Gruppioni2013,Magnelli2014}.
Here we choose to use an IR luminosity function given in ~\citet{Gruppioni2013}, which can be written as
\begin{equation}
	\Phi = \Phi_*\left(\frac{L_{\rm IR}}{L_{\rm IR}^*}\right)^{1-\alpha}\exp\left[-\frac{1}{2\sigma^2}
		\log^2\left(1+\frac{L_{\rm IR}}{L_{\rm IR}^*}\right)\right].
\end{equation}
Here the redshift evolution of parameters $\Phi_*$, $L_{\rm IR}^*$, $\alpha$, and $\sigma$ are found to be
\begin{equation*}
	\Phi_*=\left\{
	\begin{aligned}
		5.7\times10^{-3}(1+z)^{-0.57}\quad z \leq1.1 \\
		6.81\times10^{-2}(1+z)^{-3.92}\quad z>1.1    \\
	\end{aligned}
	\right .
\end{equation*}

\begin{equation*}
	L_{\rm IR}^*=\left\{
	\begin{aligned}
		7.68\times10^9(1+z)^{3.55}\quad z\leq1.85 \\
		5.80\times10^{10}(1+z)^{1.62}\quad z>1.85 \\
	\end{aligned}
	\right .
\end{equation*}

\begin{equation*}
	\alpha=\left\{
	\begin{aligned}
		1.15\quad z\leq0.3 \\
		1.2\quad z>0.3     \\
	\end{aligned}
	\right .
\end{equation*}

\begin{equation*}
	\sigma=\left\{
	\begin{aligned}
		0.52\quad z\leq0.3 \\
		0.5\quad z>0.3     \\
	\end{aligned}
	\right.
\end{equation*}

\begin{figure*}
	\centering
	\includegraphics[width=0.81\columnwidth]{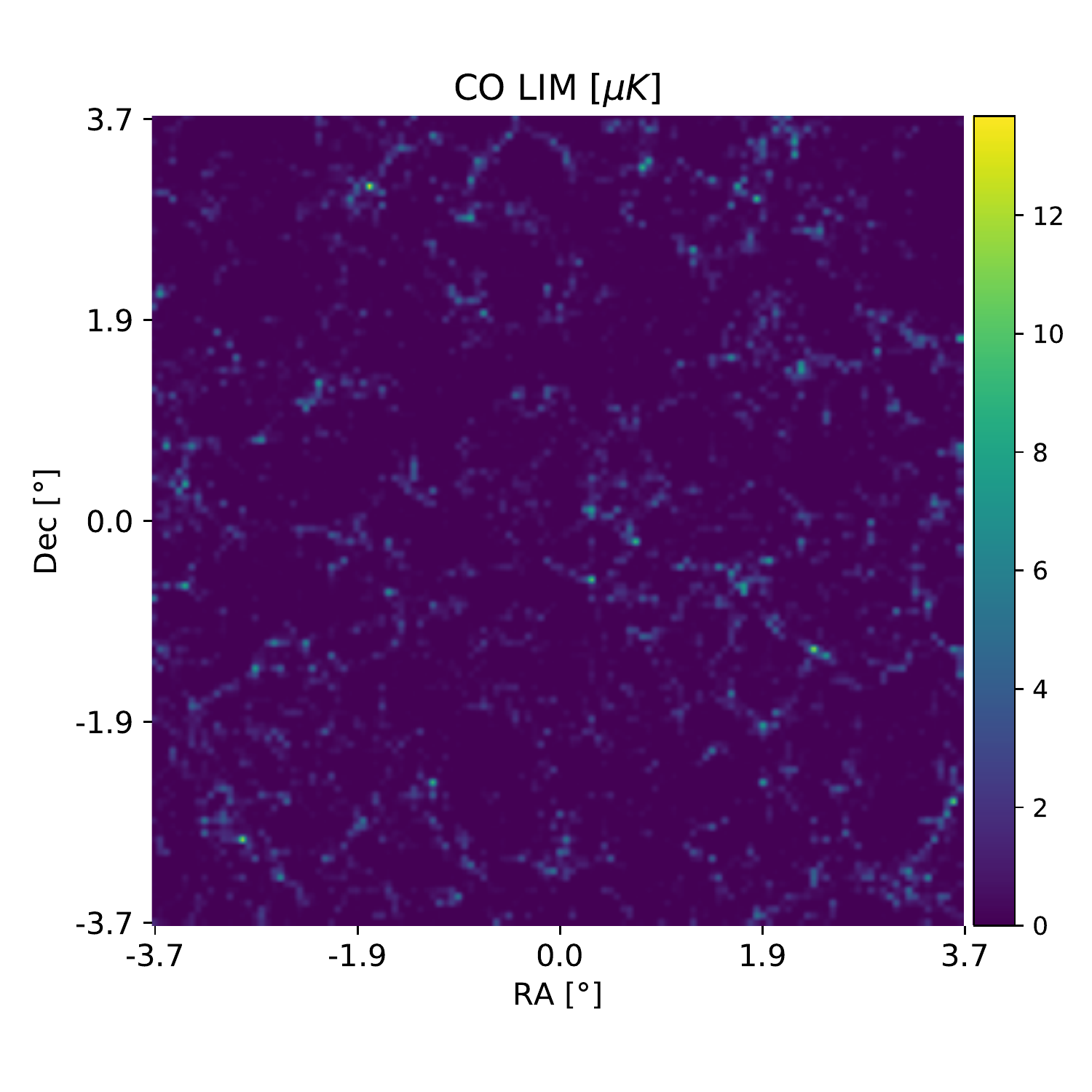}
	\includegraphics[width=1.1\columnwidth]{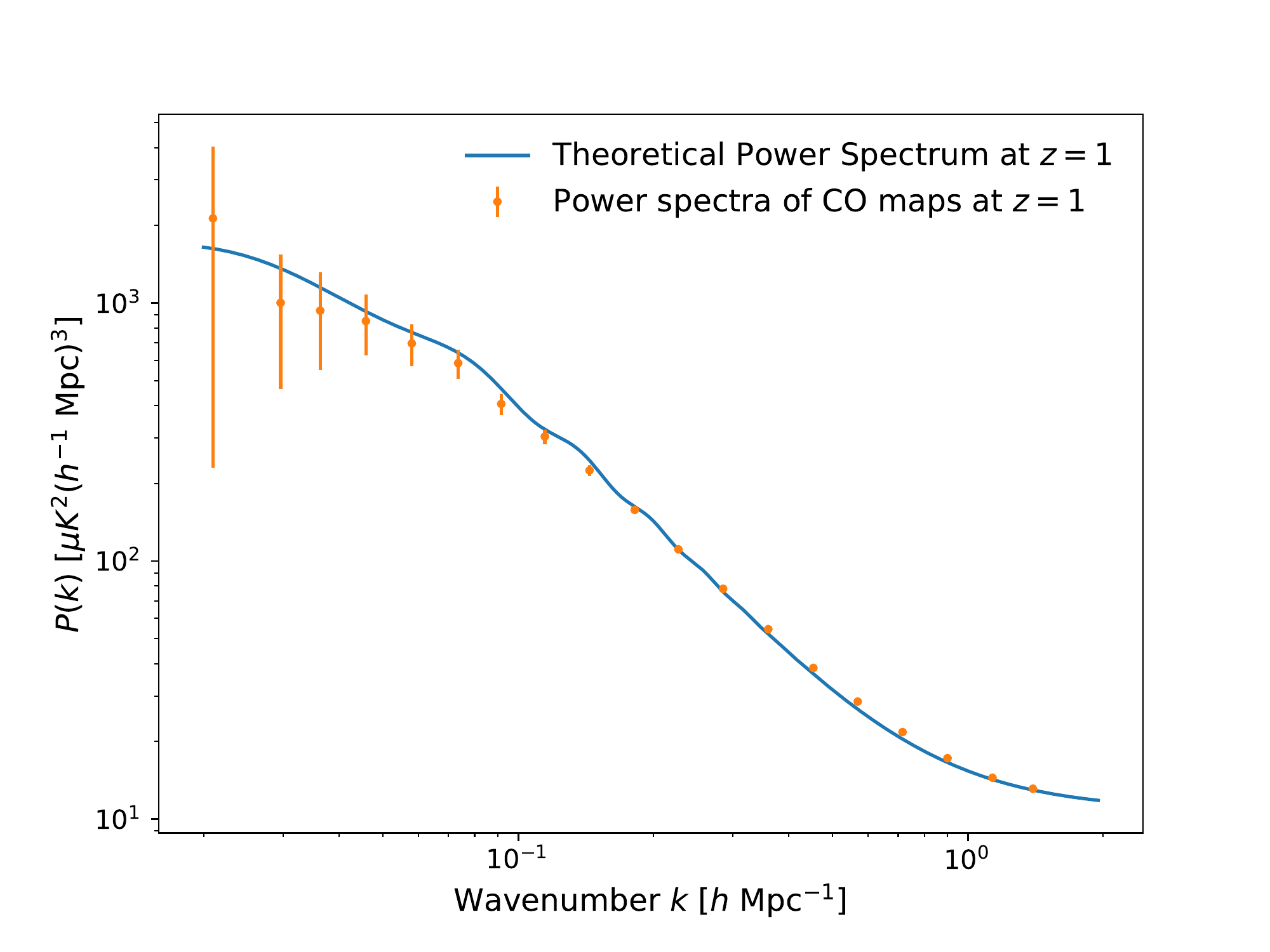}
	\caption{{\it Left panel}: an example of CO(1-0) signal mock maps at 57.6 GHz (slice of $z\sim1$).  {\it Right panel}: the corresponding CO(1-0) signal power spectrum derived from the mock map, and the theoretical power spectrum assuming the same parameter values is also shown in blue curve for comparison.}
	\label{fig:example and ps}
\end{figure*}

Considering abundance matching (Equation~\ref{eq:abundance matching}), relation between CO and IR luminosity (Equation~\ref{eq:CO and IR relation}) and unit conversion (Equation~\ref{eq:unit conversion}), the relation between CO luminosity $L_{\rm CO}$ and halo mass $M_{\rm h}$ can be derived. We show this relation for our fiducial model at $z=1$ in Figure~\ref{fig:CO_Hm relation}. The uncertainty of this relation is also shown in orange band, which is derived from the errors of $\alpha$ and $\beta$. This result is also consistent with other studies \citep[e.g.][]{Padmanabhan2018}. After obtaining $L_{\rm CO}-M_{\rm h}$ relation, we can derive the CO line intensity given by Equation~\ref{eq:observed CO intensity}, and generate the CO signal intensity maps.

Here we take spatial resolution as $1024\times1024$ with angular size $\Delta\theta$ of each pixel as $0.44$ arcmin. Note that this large resolution is prepared for adding beam effect that will be discussed later.
Then each CO intensity map is a 3 dimensional matrix with size $1024\times1024\times51$, and the third dimension indicates the number of frequency slices. Considering the 6 different directions of line-of-sight (LOS) and 27 subboxes cut from the whole simulation box, we obtain 162 CO signal intensity maps in total. We show an example of CO intensity maps downsampled to $128\times128$ at 57.6 GHz (slice at $z\sim1$) in the left panel of Figure~\ref{fig:example and ps}. We also compare the derived line power spectrum from the mock map with the theoretical one using the same parameter values in the right panel \citep{Gong2011}. The errorbars of the line power spectrum are calculated by the root-mean-square of errors from power spectrum of each map. We can see that the cosmic large-scale structure can be clearly recognized in the CO map, and the power spectrum of the simulated CO map is in good agreement with the theoretical one.

\subsection{Foregrounds}
\label{sec:foreground}

\begin{figure*}
	\includegraphics[width=2\columnwidth]{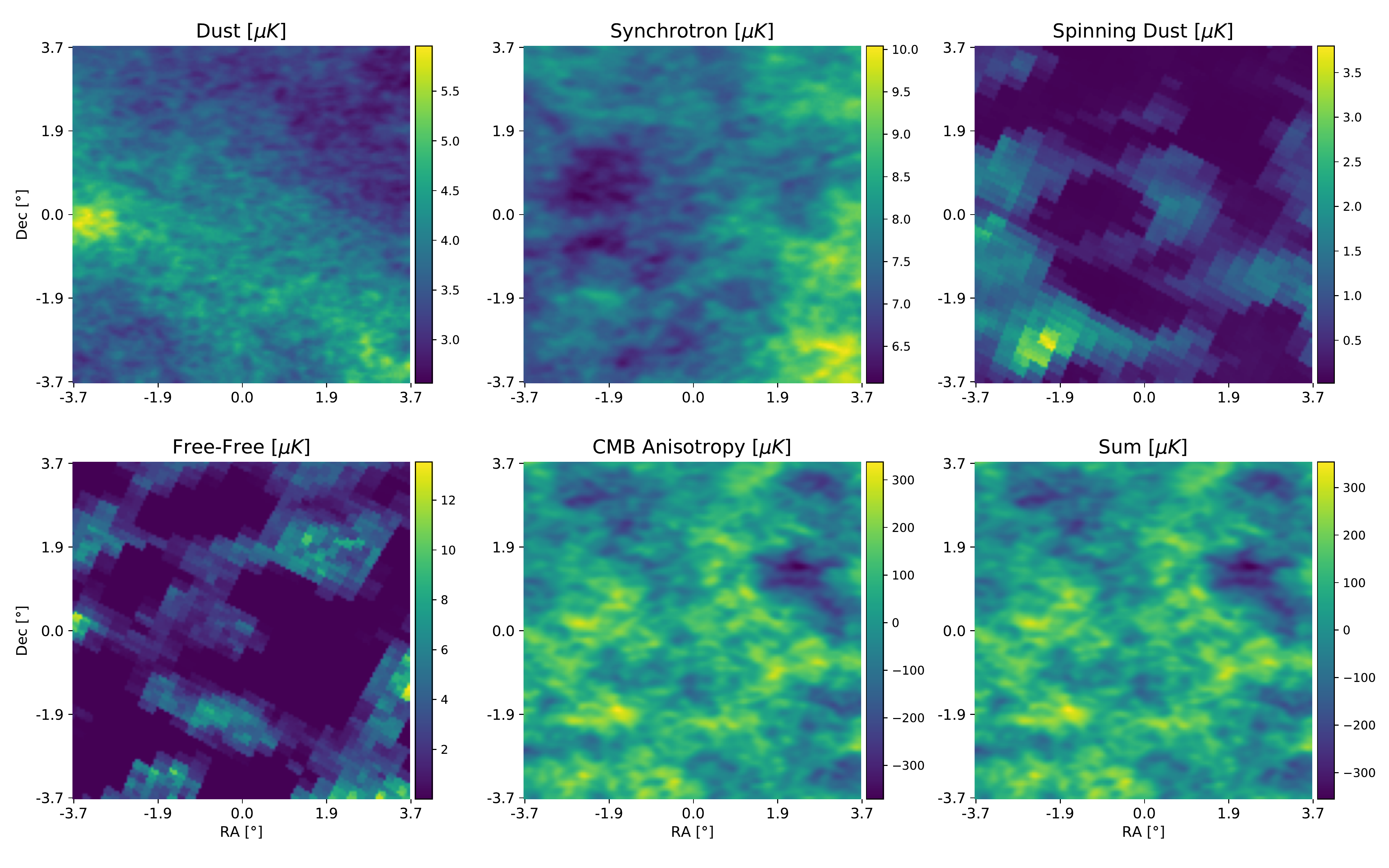}
	\caption{The foreground components at 57.6 GHz, including thermal dust, synchrotron, spinning dust, free-free emissions, CMB anisotropy and the sum of them. They are sliced by healpix with similar angular resolution of the CO signal maps. We notice that the spinning dust and free-free emissions have some grid patterns, which are caused by the extrapolation of angular power spectra to small scales.}
	\label{fig:foregrounds}
\end{figure*}

From the brightness temperature rms of astrophysical components with respect to frequency in Planck 2015 results~\citep{Adam2016}, we find that our CO maps within about $55\sim60$ GHz can be contaminated by several foreground components, including thermal dust, spinning dust, synchrotron, free-free emissions and CMB anisotropy.
We use PySM3 python package~\citep{Thorne2017} to generate all these foreground components. PySM3 provides an easy-to-use interface to simulate the sky based on publicly available data from WMAP and Planck surveys. The sky components are extrapolated to user-specified frequencies using scaling law and maps of spectral parameters. And the high resolution sky can be obtained through extrapolating the angular power spectrum of available data to smaller scales. We obtain the sky maps of foreground components using this package with nside as 8192. This resolution corresponds to an angle of $0.43$ arcmin, which is approximately equivalent to the pixel resolution 0.44 arcmin of the CO signal maps. Therefore, the slicing of foreground can be performed without smoothing the information at small scale, and this resolution is high enough for simulating the beam effect in the survey.

After obtaining the sky maps of the foreground components, we generate square maps using healpix with the same spatial pixels as the CO intensity maps in declinations higher than 50 degree and lower than $-50$ degree. This criteria is to ensure that the foreground emissions are not too large compared to that around galactic disc. For sky area of $7.45\times7.45\deg^2$ in the simulation, we can obtain 48 foreground maps surrounding one circle of the whole sky and in total 96 maps for northern and southern hemispheres at frequencies the same as the CO maps. Each foreground component and the sum of them at 57.6 GHz are shown in Figure~\ref{fig:foregrounds}. We notice that the spinning dust and free-free emissions have some grid effects, and they are caused by extrapolation of angular power spectra to small scales.

Note that we do not consider the contamination from other emission lines, such as CO(2-1) and CO(3-2) that can contaminate the CO(1-0) signal from higher redshifts. Since CO(1-0) line is luminous and has a long wavelength, the main contamination from other lines usually comes from higher redshifts, and these lines are then redshifted and become relatively not very strong. On the other hand, the contamination from interloper lines also can be effectively distinguished by the shape difference from the signal line in a redshift-space line intensity power spectrum \citep{Visbal2010,Gong2014,Lidz2016}, or can be modeled as cosmological signals from other redshifts \citep{Gong2020}.  For simplicity, we ignore the contamination from these interloper lines, and only focus on the foregrounds mentioned above.

\subsection{Instrumental effects}
\label{sec:instrumental effects}

\begin{table}
	\caption{Instrumental parameters considered in this work.}
	\begin{tabular}{lc}
		\hline
		Parameter                                        & Value            \\ \hline\hline
		Telescope aperture, $D$ [m]                      & 10.4             \\
		Survey area, $A_{\rm S}$ [$\deg^2$]              & 7.45$\times$7.45 \\
		Total observing time, $\tau_{\rm tot}$ [hrs]     & 9000             \\
		Beam size at 57.6 GHz, $\theta_{\rm d}$ [arcmin] & 1.72             \\
		Number of feeds, $N_{\rm feed}$                  & 100              \\
		Frequency range, $B$ [GHz]                       & 55-60            \\
		Frequency resolution, $\Delta\nu$ [GHz]          & 0.1              \\
		System temperature, $T_{\rm sys}$ [K]            & 40               \\ \hline
	\end{tabular}
	\label{tab:instrumental parameters}
\end{table}

The instrumental effects we consider are white noise and beam effects of radio telescope. Here we derive the white noise from an assumed  instrument with some of parameters similar to the CO Mapping Array Project (COMAP)~\citep{Cleary2022}. The instrumental noise per voxel can be estimated by
\begin{equation}
	\label{eq:noise}
	\sigma_T = \frac{T_{\rm sys}}{\sqrt{\tau\Delta\nu}} = \frac{T_{\rm sys}
	\sqrt{N_{\rm pixel}}}{\sqrt{\tau_{\rm tot}e_{\rm obs}N_{\rm feed}\Delta\nu}}.
\end{equation}
We assume the total observing time $\tau_{\rm tot}=9000$ hrs, efficiency of observation $e_{\rm obs}=1$, $N_{\rm feeds}=100$,  the frequency resolution $\Delta \nu=0.1$ GHz, and the system temperature $T_{\rm sys} = 40{\rm K}$. The number of pixels $N_{\rm pixel}$ can be estimated by the survey area and beam size, and we find $N_{\rm pixel}=260\times260$ at $z=1$. The instrumental parameters are summarized in Table~\ref{tab:instrumental parameters}. Then we calculate the instrumental noise per voxel and find that $\sigma_T=18.3\,{\rm \mu K}$ at 57.6 GHz. The noise is evenly added over voxels of the CO maps by Gaussian distribution. In the following discussion, we will firstly ignore the instrumental noise, and then discuss its effect on foreground removal.

\begin{figure*}
	\centering
	\includegraphics[width=\columnwidth]{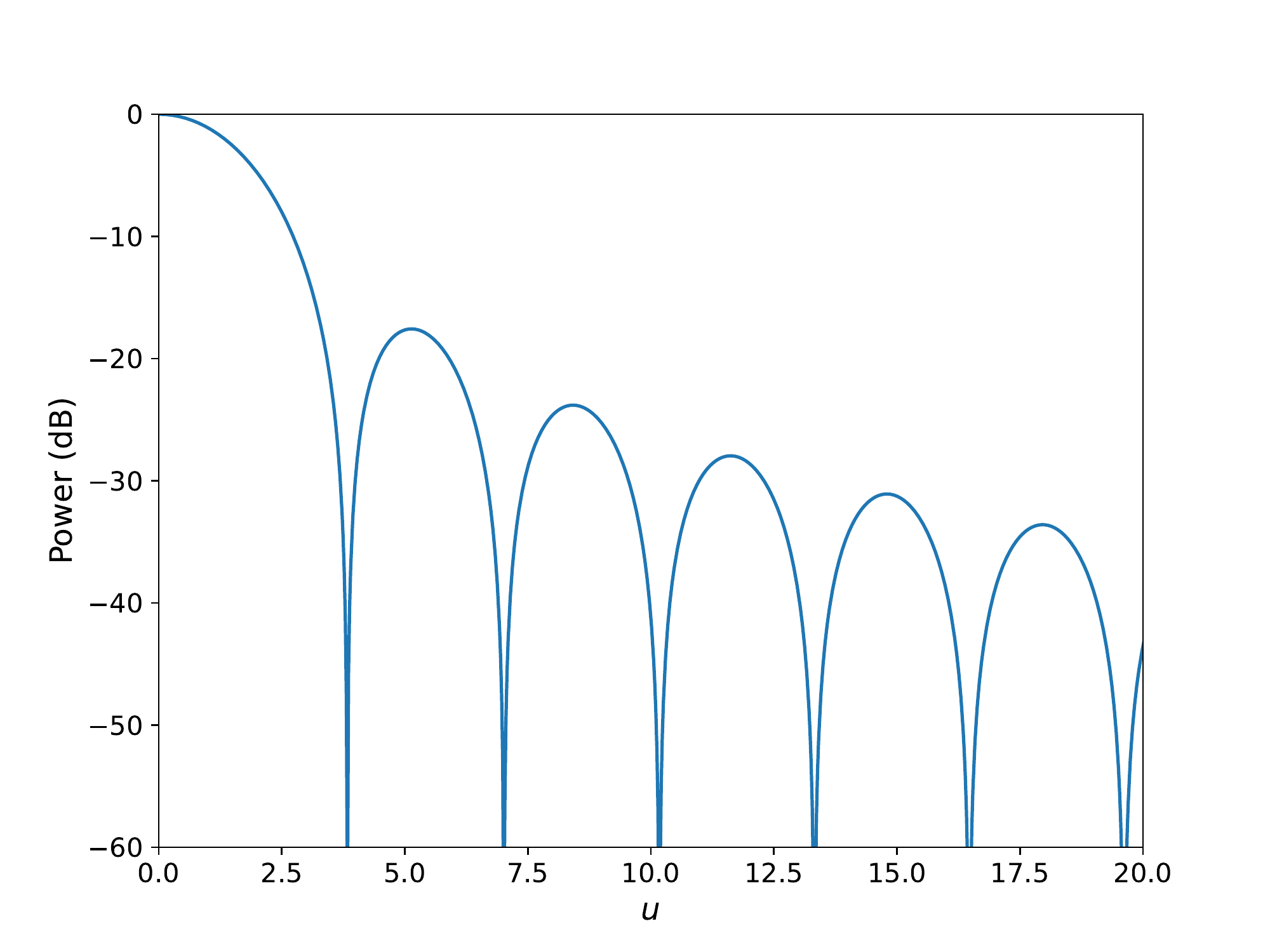}
	\includegraphics[width=\columnwidth]{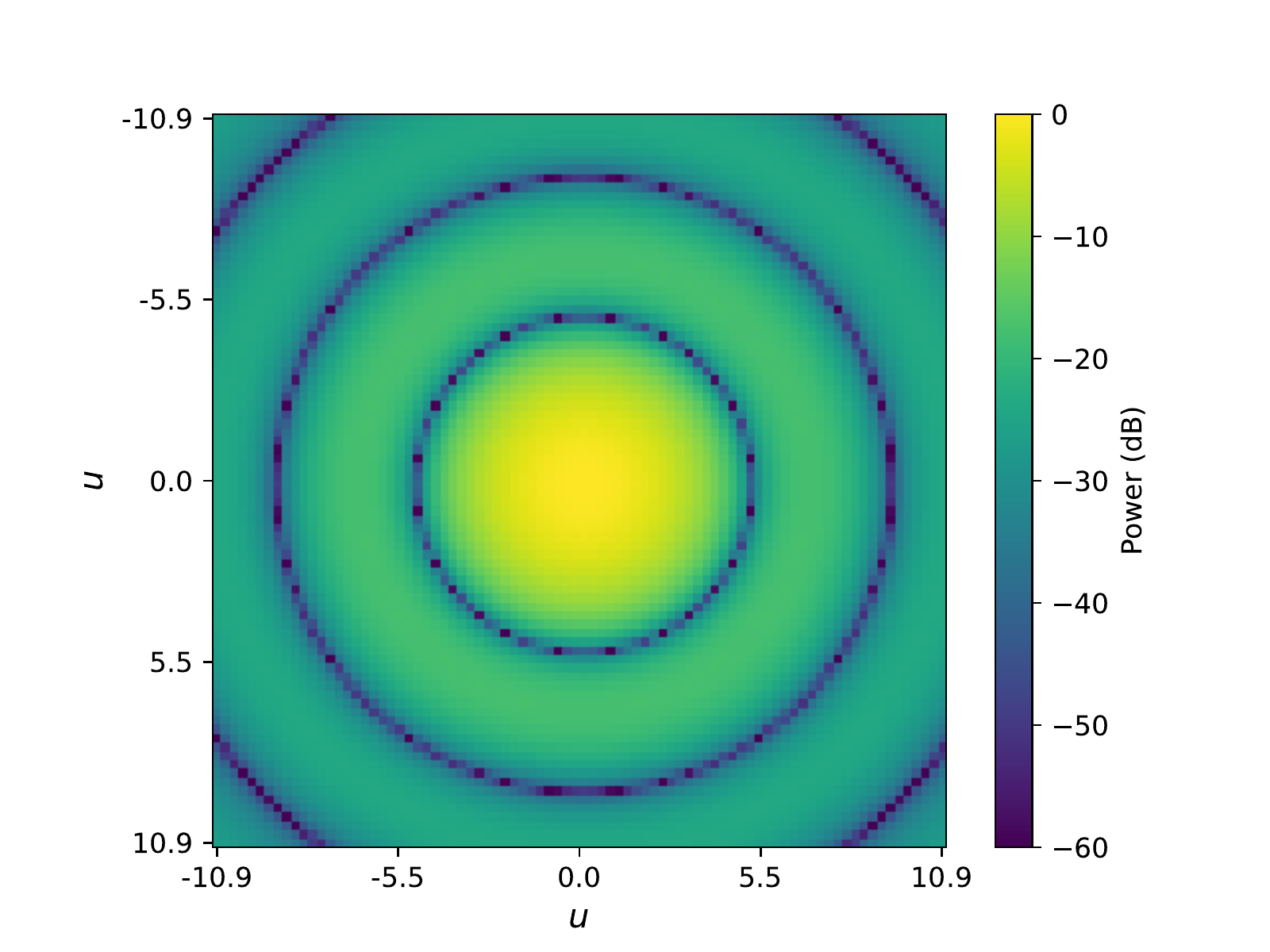}
	\caption{The Bessel model (left panel) and the 2 dimensional Bessel kernel (right panel) at 57.6 GHz we assume in this work. }
	\label{fig:beam model}
\end{figure*}

The beam characterizes the power response of a radio telescope to the sky signal, and the observed sky map is the convolution of the real sky and the beam of telescope. The beam of a telescope is generally chromatic, i.e. the shape of beam will change with respect to frequency. The chromaticity will cause fluctuation in the spectrum of foreground and increase the degeneracy between signal and foreground, and therefore potentially cause more signal loss or foreground leakage in foreground removal.

Gaussian function is widely used to model the beam. However, this simplified model lacks sidelobe and therefore underestimate the chromaticity. Other beam model with sidelobe of different level is proposed, e.g. Cosine model~\citep{Condon2016}, gaussian tapered Airy dish~\citep{Harper2018} and Bessel or Jinc beam~\citep{Wilson2013,Matshawule2021}, based on different aperture tapering. Beam models based on electromagnetic simulation and astro-holographic observation are also proposed~\citep{Asad2021,Sun2022,Matshawule2021}.
Here we use the Bessel model corresponding to uniformly illuminated circular aperture. We notify that the Bessel model is the worst case of sidelobe effect, because any tapering toward the edge of aperture will generally suppress the level of sidelobe.

The power response of the Bessel model can be written as
\begin{equation}
	\label{eq:beam model}
	F(\sin{\theta}) = \left(2\frac{J_1(u)}{u}\right)^2,
\end{equation}
where
\begin{equation*}
	u = \frac{\sin{\theta}}{(\theta_{\rm d}/\pi)}\ {\rm and}\ \theta_{\rm d} = \frac{\lambda}{D}.
\end{equation*}
Here $\lambda$ is the observed wavelength, $D$ is the aperture of the antenna assumed to be $10.4\,\rm m$ the same as COMAP, and $J_1$ is the Bessel function of the first kind.
We plot the Bessel model and 2 dimensional Bessel kernel at 57.6 GHz in Figure~\ref{fig:beam model}. We notice that the first
sidelobe is at $u\sim5$ or 2.86 arcmin, which is larger than the pixel resolutions of CO maps and foregrounds, and thus the sidelobe effects can be well simulated. We construct a beam kernel with the same resolution as the CO and foreground maps by including the first sidelobe, and convolve with the CO intensity maps and foregrounds to simulate the beam effect. Then we downsample these maps to spatial resolution of $128\times128$ by averaging adjacent 8 pixels, and the instrumental noise level becomes $\sigma=9.0\ {\rm \mu K}$ per voxel. The spatial pixels as powers of 2 are easy for building and training of neural networks.
We also notice that the peculiar velocity can degenerate with the beam effect, that will suppress the line power spectrum at small scales. Since the intensity mapping mainly explores the large scales and this work is focusing on the foreground removal, for simplicity, we ignore the peculiar velocity and other redshift-space distortion effect when generating the mock maps.

\section{Deep learning method}
\label{sec:methods}

We use Principal Component Analysis (PCA) as benchmark model to remove foregrounds and compare it with our deep learning method. PCA decomposes multivariate dataset to the components which have largest variances~\citep{Wold1987, Abdi2010}. These components are called principal components (PCs) or PCA modes, corresponding to eigenvectors of covariance matrix of the data. The 2 dimensional Gaussian distributions can be decomposed to 2 PCs forming an ellipsoid for example. PCA subtraction finds the largest eigenvalues of covariance of data with largest variances and subtracts these components. This method tends to subtract foreground along with some signal information. And since it depends on the assumption that the foregrounds are smooth across frequencies, the foreground removal is further complicated considering the beam effect that is also correlated with frequencies. These two effects will lead to deviations of power spectra with true powers. Deep learning algorithm concerning image-to-image generation may recover these deviations by learning features from images. Our networks are implemented by Keras~\footnote{https://keras.io} with TensorFlow as backend~\footnote{https://www.tensorflow.org/}.

\renewcommand{\arraystretch}{1.5}
\begin{table}
	\caption{Details of ResUNet architecture.}
	\label{tab:resunet}
	\begin{center}
		\begin{tabular}{|c|c|c|}
			\hline
			Layers                                 & Output Status$^a$ & Number of params.$^b$ \\ \hline
			\hline
			Input                                  & (128, 128, 51)    & 0                     \\ \hline
			Conv2d                                 & (128, 128, 64)    & 29,440                \\ \hline
			LeakyReLU \large{\textcircled{1}}      & (128, 128, 64)    & 0                     \\ \hline
			ResBlock \large{\textcircled{2}}$^c$   & (64, 64, 128)     & 130,656               \\ \hline
			ResBlock \large{\textcircled{3}}       & (32, 32, 256)     & 519,360               \\ \hline
			ResBlock \large{\textcircled{4}}       & (16, 16, 512)     & 2,070,912             \\ \hline
			ResBlock \large{\textcircled{5}}       & (8, 8, 512)       & 8,270,592             \\ \hline
			ResBlock \large{\textcircled{6}}       & (4, 4, 1024)      & 8,928,000             \\ \hline
			ResBlock + \large{\textcircled{5}}$^d$ & (8, 8, 1536)      & 11,290,112            \\ \hline
			ResBlock + \large{\textcircled{4}}     & (16, 16, 1024)    & 9,192,448             \\ \hline
			ResBlock + \large{\textcircled{3}}     & (32, 32, 512)     & 2,434,560             \\ \hline
			ResBlock + \large{\textcircled{2}}     & (64, 64, 256)     & 611,072               \\ \hline
			ResBlock + \large{\textcircled{1}}     & (128, 128, 128)   & 153,984               \\ \hline
			Conv2d                                 & (128, 128, 51)    & 58,803                \\ \hline
			Square                                 & (128, 128, 51)    & 0                     \\ \hline\hline
			\\
		\end{tabular}
	\end{center}
	\vspace{-2mm}
	\textbf{Notes}.\\
	$^a$ Format: (dimension, dimension, channel).\\
	$^b$ Total number of parameters: 43,689,939.\\
	$^c$ ResNet1 + ResNet2.\\
	$^d$ ResNet1 + concatenation + ResNet2.\\
\end{table}

\begin{figure*}
	\includegraphics[width=2\columnwidth]{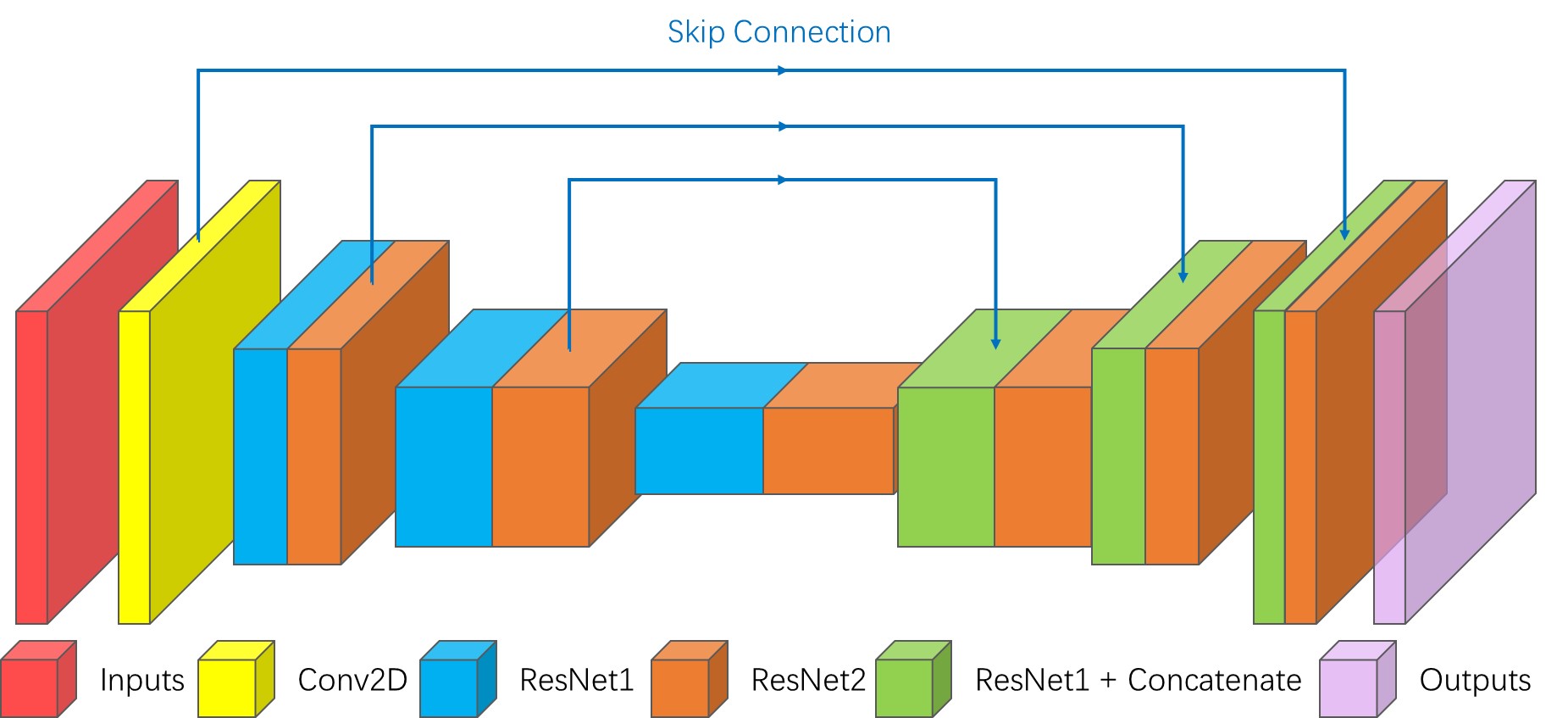}
	\caption{Schematic diagram of ResUNet. The ResNet1 and ResNet2 block are built crossly. The skip connection concatenates
		features learned in encoding path with the ones in decoding path.}
	\label{fig:resunet}
\end{figure*}

\subsection{ResUNet}
\label{sec:unet}
\begin{figure}
	\includegraphics[width=\columnwidth]{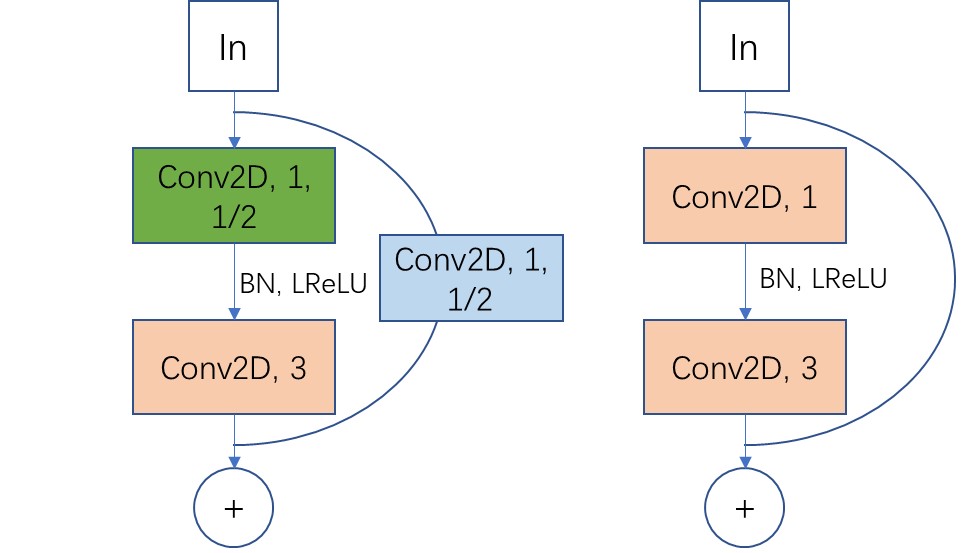}
	\caption{ResNet1 (left) and ResNet2 (right) used in our work. ResNet1 downsamples the spatial dimension of feature maps by 2 using strided convolution, and the inputs are also downsampled and added to the output of the main convolutional layer. ResNet2 will keep the spatial dimension. BatchNormalization and LeakyReLU are employed to reduce overfitting and apply non-linearity after convolution respectively.}
	\label{fig:resnet}
\end{figure}

Our deep learning model is built upon ResUNet~\citep{Diakogiannis2020}, which effectively combines UNet with ResNet. UNet is introduced in biomedical image segmentation~\citep{Ronneberger2015}, and is based on convolutional neural network (CNN). To perform image-to-image generation task, UNet uses a bottleneck structure. It convolves and downsamples the maps to learn the relevant features in encoding path. In decoding path where it generates images, UNet uses transposed convolution or upsampling-convolution to increase the spatial dimensions of feature maps to output size. And it uses a bridge to connect the features learned from encoding path with decoding path to increase its generative ability.

Our UNet uses modified structure called ResUNet~\citep{Diakogiannis2020}, which is constructed based on ResNet~\citep{He2015}.
ResNet is a milestone in development of deep learning field, and it helps tackling the vanishing gradient problem using identity mapping method. Since convolutional neural network increases its ability by stacking more convolutional layers, we can build very deep networks using ResNet block without worrying the training problem induced by vanishing gradient problem. Two types of ResNet blocks are employed in our work, called ResNet1 and ResNet2. Using a scaling layer, ResNet1 can downsample the spatial dimension of maps by a factor of 2, and the inputs are also downsampled and added to the outputs from the main convolution path. On the other hand, ResNet2 keeps the spatial dimension using the same-padding convolutional layers. In decoding path, ResNet1 is altered to upsample the spatial dimension. \citet{He2015} employs two convolutional layers using same kernel sizes in his originally implemented ResNet block, however, here we alter the kernel size of first layer to be $1\times1$ for decreasing the channels and reduce the computing expense. We have proven that this modified structure has no influence on performance. The upsampling can be performed by UpSampling or transposed convolutional layers. Here we choose UpSampling to avoid the potential checkerboard effect~\citep{Odena2016}. BatchNormalization~\citep{Ioffe2015} and LeakyReLU~\citep{Maas2013} are employed to reduce overfitting and apply non-linearity after convolution, respectively. LeakyReLU is improved from ReLU and allows a small gradient when some units are below 0, and the negative slope parameter of LeakyReLU is taken to be 0.3.

The inputs are firstly convolved by a same-padding convolutional layer with 64 kernels of size 3, obtaining a feature map with the
same spatial size as inputs. After convolution, we use LeakyReLU as our activation function. Next we stack 5 ResNet1 and ResNet2 blocks crossly. The feature maps are downsampled by 2 every time they pass ResNet1, and ResNet2 is employed to increase the learning ability by stacking more convolutional layers. Finally, the inputs are encoded in 3d feature matrix of $4\times4\times1024$. The decoding path is
almost mirror of encoding path. We upsample the feature maps by ResNet1 with UpSampling, concatenate with maps obtained in encoding path, and further process them with ResNet2. After upsampling 5 times, maps reach the same spatial dimension as inputs. We use a convolutional layer to scale the channels to 51 to obtain the outputs with the same size of inputs. Finally we apply square function to the outputs to restrict positive maps since the CO signals are all positive. The schematic diagrams of ResNet block and our ResUNet are displayed in Figure~\ref{fig:resnet} and Figure~\ref{fig:resunet}, respectively. The details of our ResUNet architecture can be found in Table~\ref{tab:resunet}.

\begin{figure*}
	\includegraphics[width=2\columnwidth]{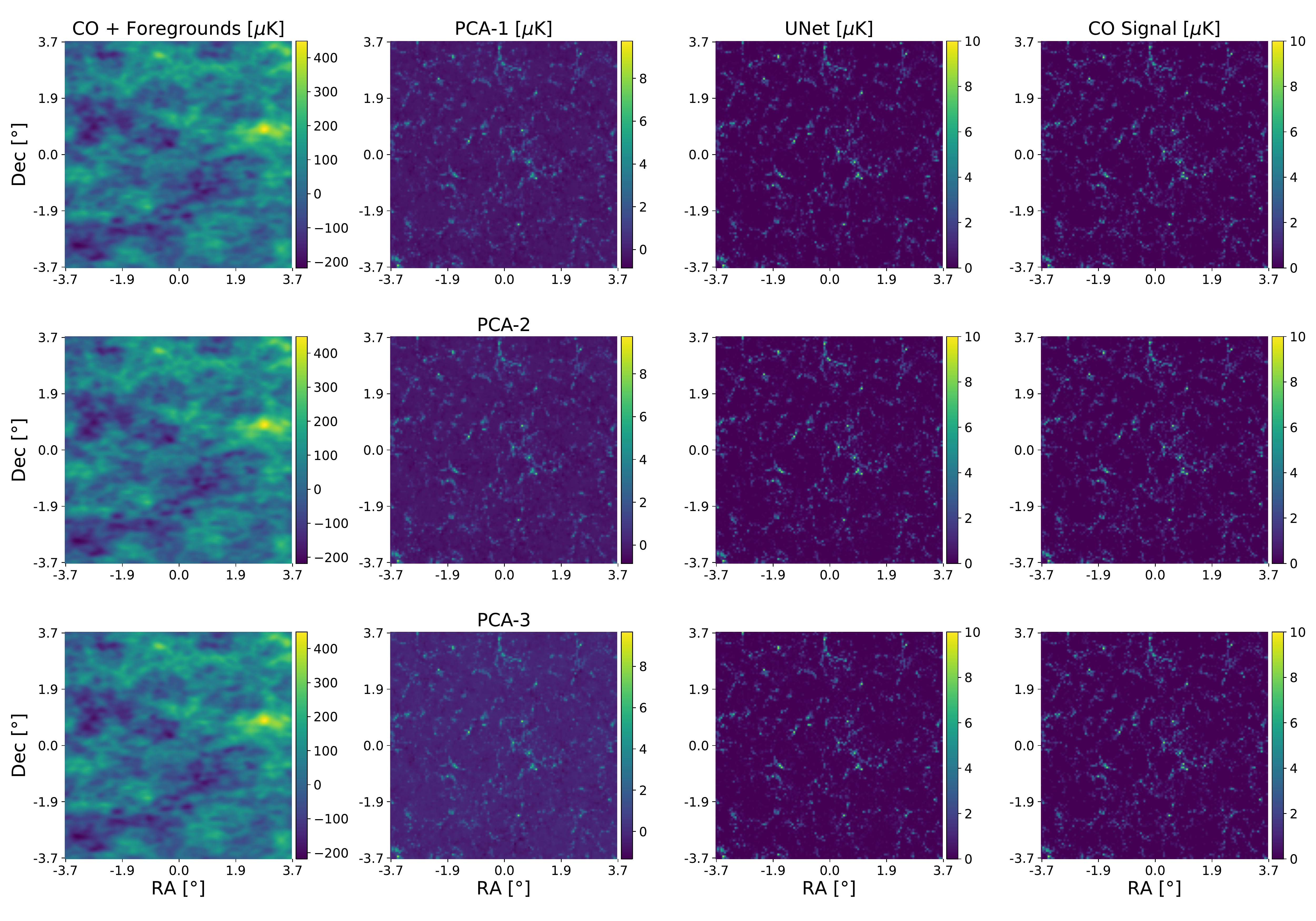}
	\caption{The CO foreground contaminated, PCA preprocessed, UNet generated, and signal (or target) intensity maps for our fiducial CO model at 57.6 GHz (slice at $z\sim1$). The first, second, and third rows denote the cases of removing the first one (PCA-1), two (PCA-2), and three  (PCA-3) largest modes in the PCA procedure, respectively. The maps are truncated at 10 $\mu$K for better comparison. We can see that our UNet can excellently recover the web structure of CO signal map, no matter how many modes are subtracted in PCA. }
	\label{fig:map comparison fiducial}
\end{figure*}

\begin{figure*}
	\centering
	\includegraphics[width=0.66\columnwidth]{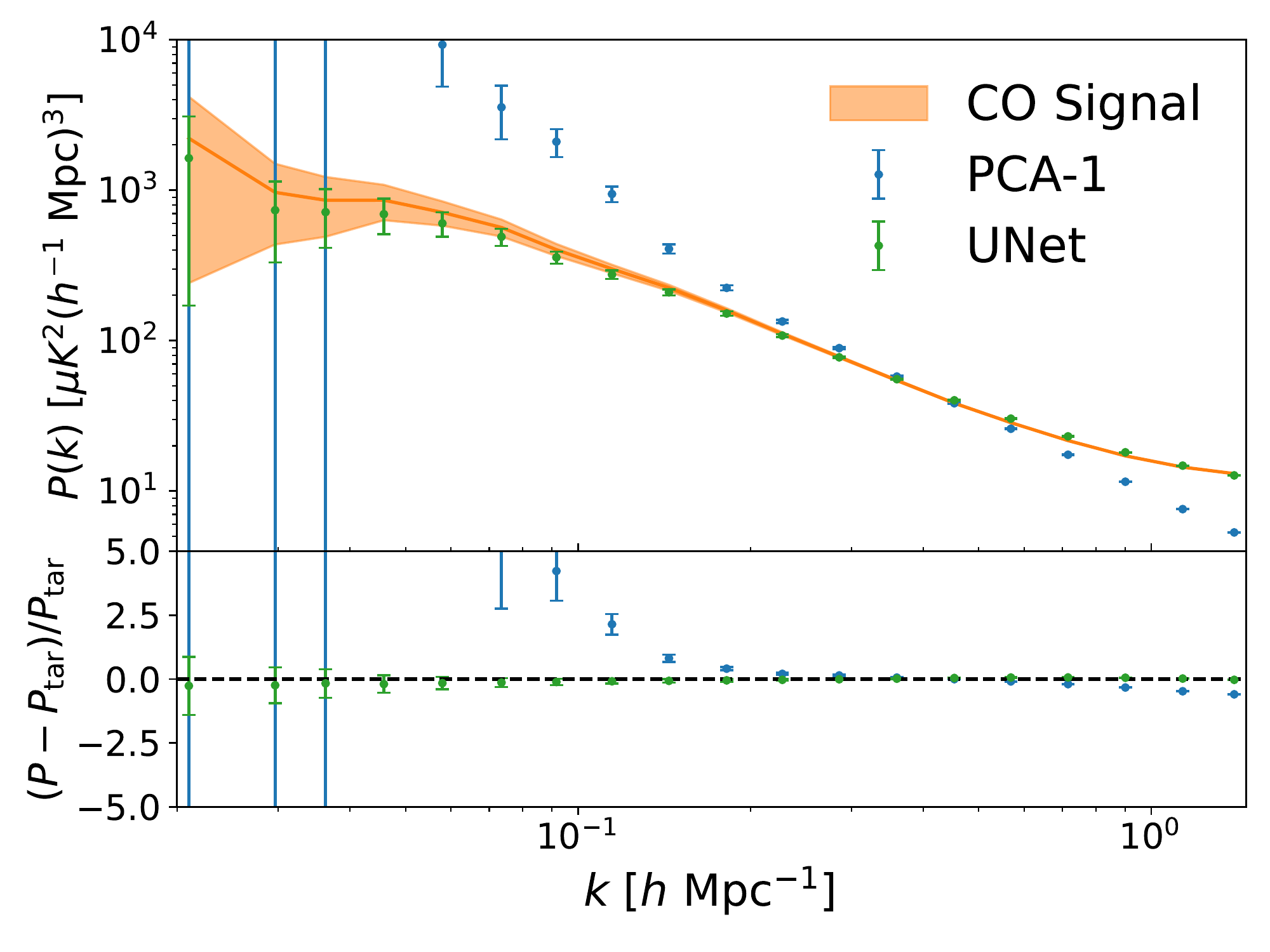}
	\includegraphics[width=0.66\columnwidth]{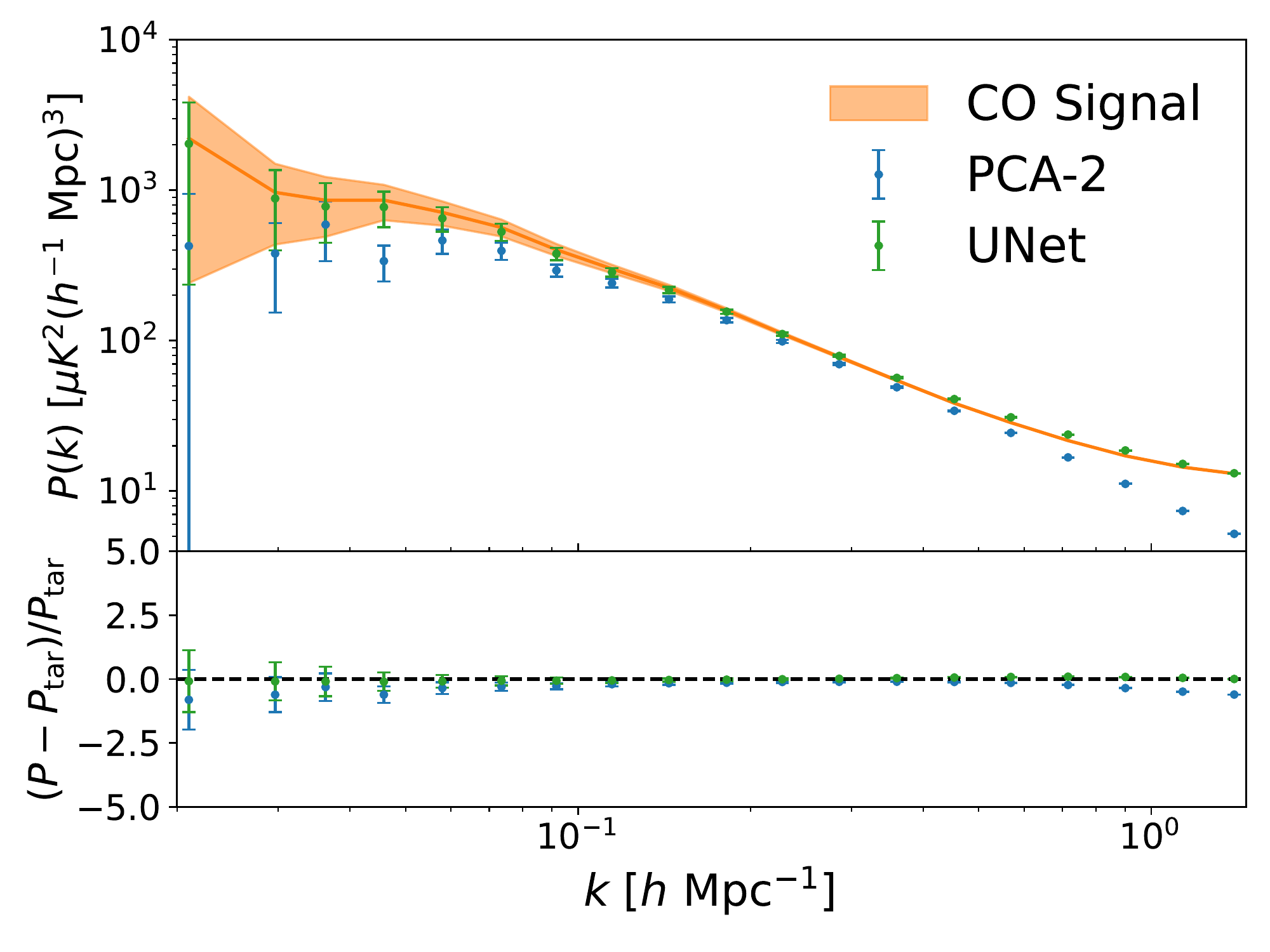}
	\includegraphics[width=0.66\columnwidth]{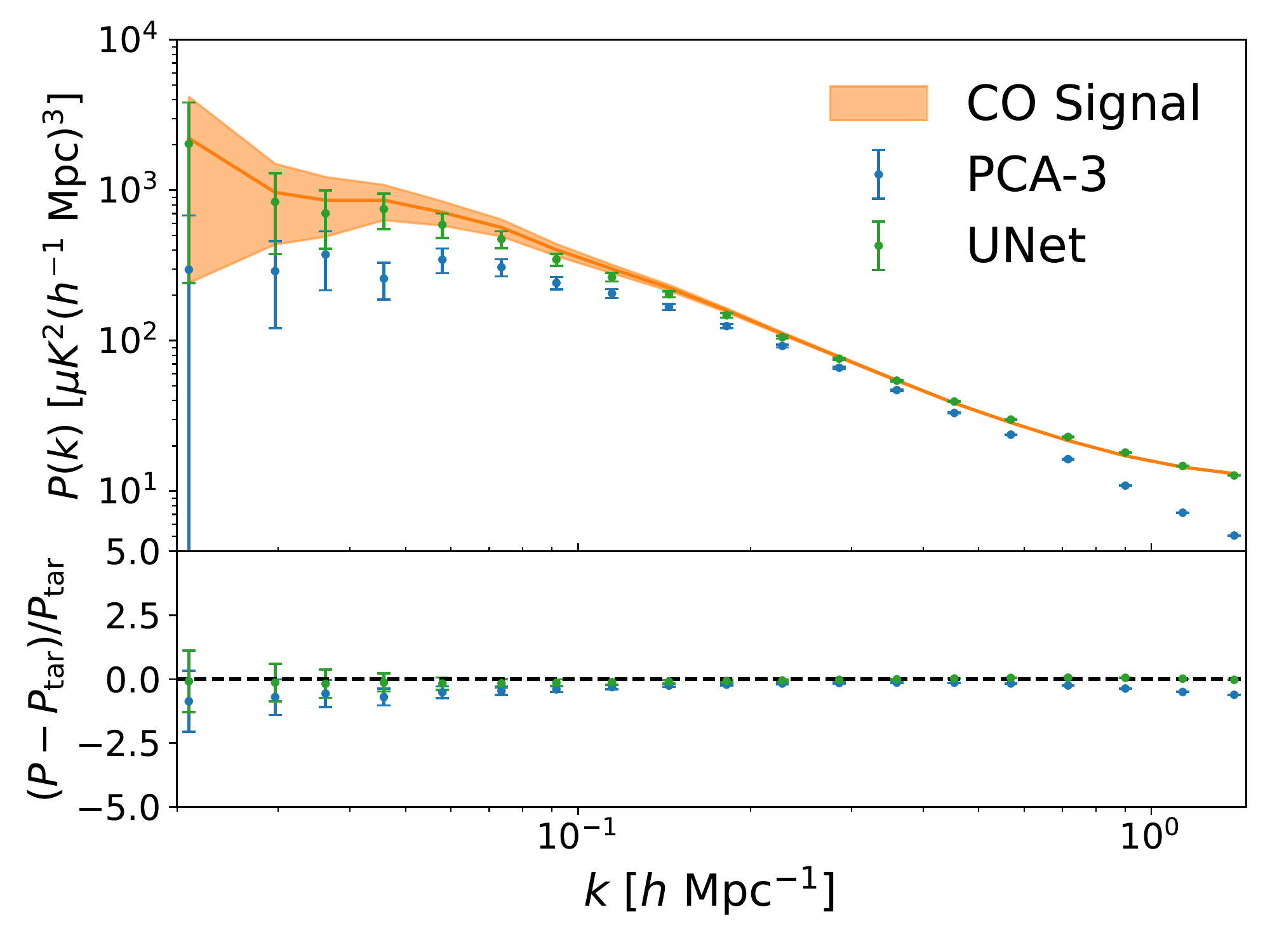}
	\caption{The CO power spectra of the PCA preprocessed, UNet generated and signal or target maps for our fiducial CO model. The errorbars are calculated by root-mean-square of all power spectra of the testing maps. The relative difference of the power spectra $(P-P_{\rm tar})/P_{\rm tar}$ is also shown in the lower panels. We notice that PCA-1 cannot completely remove the foregrounds, resulting in large deviations compared to the signal power spectrum. PCA-2 and PCA-3 can effectively clean the foregrounds but with a loss of information at both large and small (due to beam effect) scales. Our UNet can recover the power spectra excellently by correcting the residual foregrounds and beam effect, regardless of the number of modes PCA subtracts, and the derived power spectra are consistent with the CO signal power spectrum within 1$\sigma$ confidence level.}
	\label{fig:power comparison fiducial}
\end{figure*}

\begin{figure*}
	\includegraphics[width=2\columnwidth]{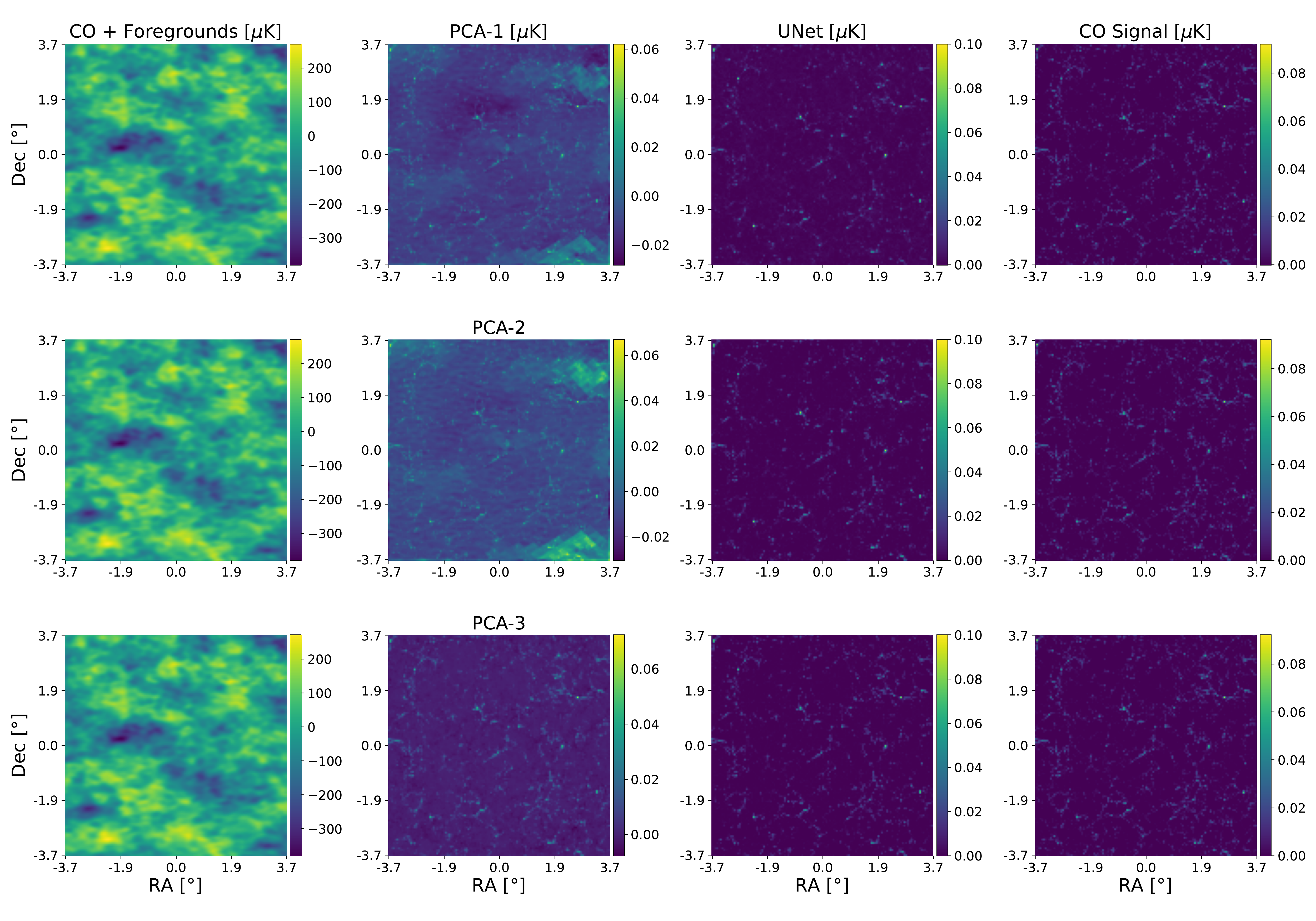}
	\caption{Same as Figure~\ref{fig:map comparison fiducial} but for our low CO signal model with the signal approximately 100 times lower compared to the fiducial one. We can see that more modes need to be removed in the PCA foreground removal, and our UNet can still correctly recover the CO signal map in different PCA mode-removing cases.}
	\label{fig:map comparison low}
\end{figure*}

\begin{figure*}
	\centering
	\includegraphics[width=0.66\columnwidth]{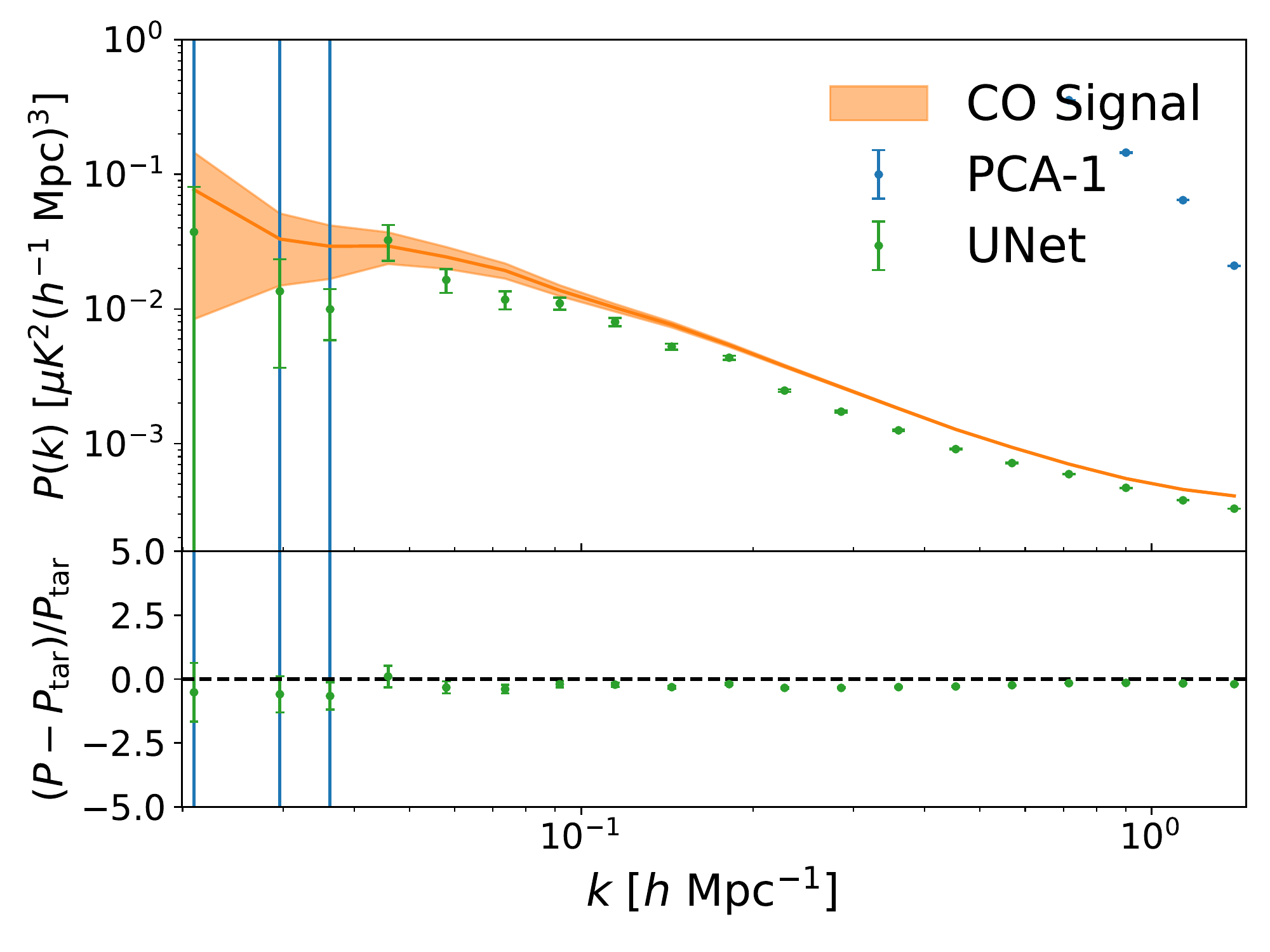}
	\includegraphics[width=0.66\columnwidth]{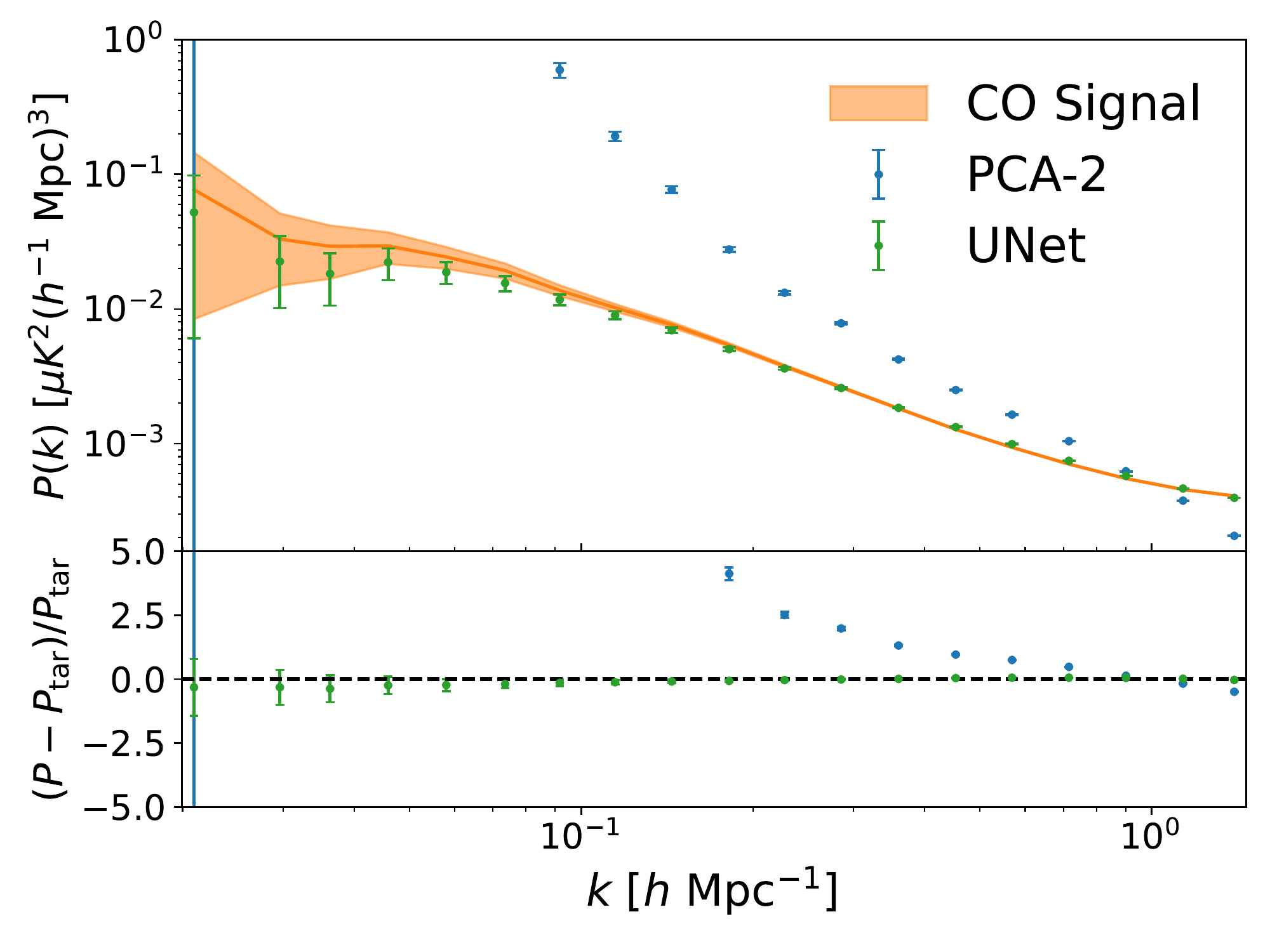}
	\includegraphics[width=0.66\columnwidth]{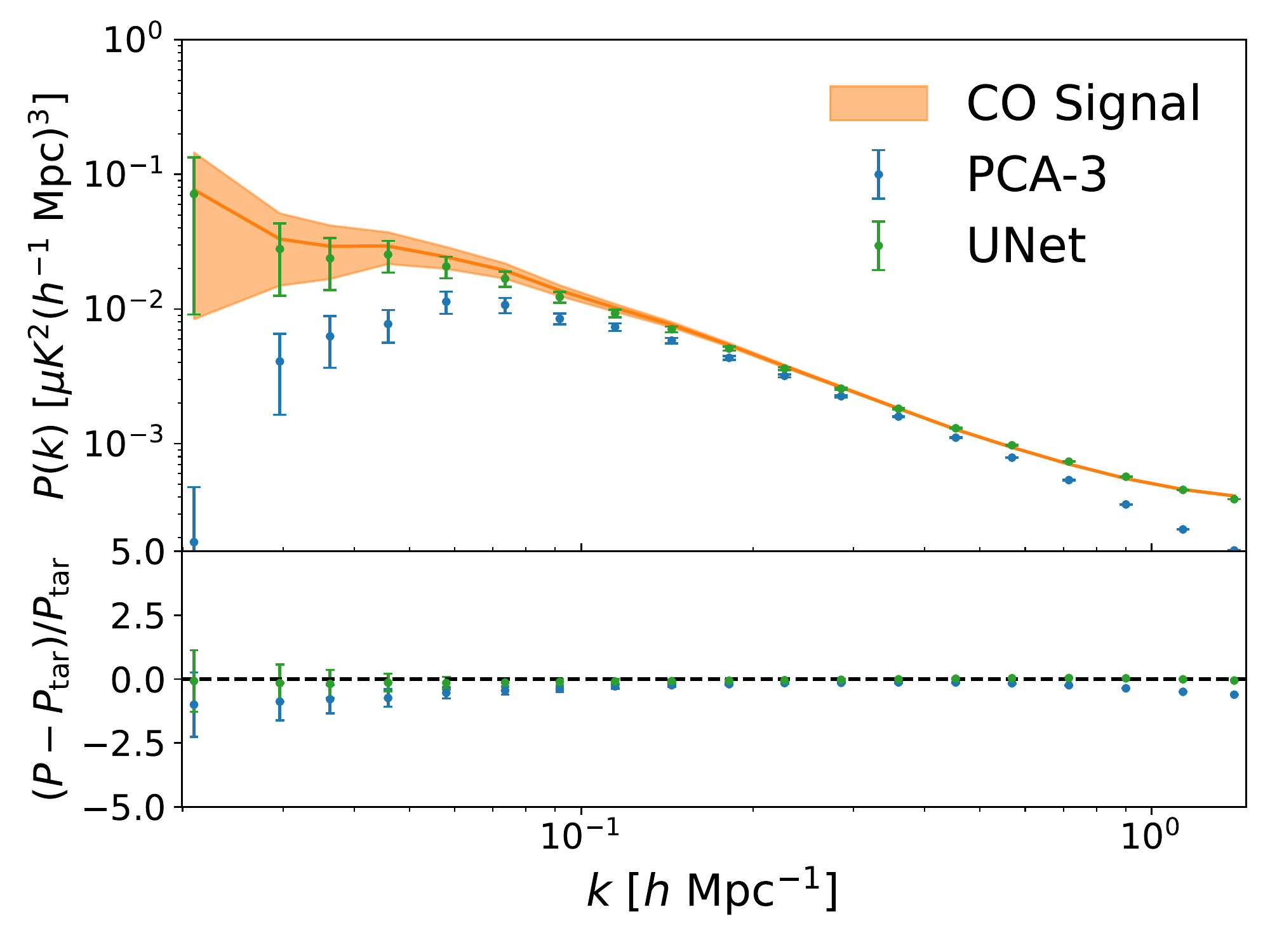}
	\caption{Same as Figure~\ref{fig:power comparison fiducial} but for our low CO signal model. There are large discrepancies between the PCA-1 or PCA-2 (blue data points) and the CO signal power spectra (orange curve) in this case. Our UNet cannot accurately obtain the CO power spectrum in the PCA-1 case, and some deviations still exist. However, the CO signal power spectra can be reconstructed within 1$\sigma$ error in the PCA-2 and PCA-3 cases, and the signal loss effect also can be effectively corrected.}
	\label{fig:power comparison low}
\end{figure*}

\subsection{Training}
\label{sec:training}

Before training, we need to split the mock data and generate the training and testing sets. We randomly select 130 of the 162 mock CO signal intensity maps and 70 of the 96 total foreground maps,  with beam effect included, and combine them together to obtain 9100 maps as the training set. The rest of them are used to create 832 mock observational maps as the testing set. Note that the number of the training and testing data are selected randomly here. As a test, we also try to reduce the number of the training data, and find that at least 2000 training data are needed to obtain a similar result.

Our inputs and targets for UNet are beam-convolved CO intensity maps contaminated by foregrounds and CO signal intensity maps without beam effect, respectively, and they are all data cubes with size $128\times128\times51$. Before feeding the UNet, these inputs are preprocessed by PCA for preliminary foreground removal and then our UNet tries to accurately recover the outputs (i.e. CO signal maps without beam effect) by removing residual foreground contamination and eliminating certain effects after PCA procedure like signal loss. We also discuss the result when inputs are not preprocessed by PCA in Section~\ref{sec:result}. We employ LogCosh as our loss function as recommended in previous studies~\citep{Makinen2021,Ni2022}. LogCosh calculates log hyperbolic cosine of deviations and is similar to the absolute function but with a smooth gradient around 0. Adam optimizer~\citep{Kingma2014} with initial learning rate $1\times10^{-4}$ is employed to optimize our network, and this learning rate can be adaptively adjusted during the training process. The batch size is 8, considering the memory limit of Graphical Processing Unit (GPU).

For low CO signal case, the amplitude of our CO signal maps are approximately 100 times lower than that in the fiducial model. To avoid extremely small gradients which can slow down the training, we scale both the PCA preprocessed inputs and targets by 100 times larger and then scale them back after training. Note that this scaling does not affect the results but only the training speed. Then the training process is fast and can reach steady losses after about 15 epochs, and we save our UNet model with the lowest loss value.

\section{Result and Discussion}
\label{sec:result}

After the training process, we feed the testing data to our UNet model for testing the network. Similar to the training process, we first preprocess the testing data with PCA for subtracting the foregrounds. Figure~\ref{fig:map comparison fiducial} shows the relevant maps at 57.6 GHz (slice of $z\simeq1$) for our CO fiducial model mentioned in Section~\ref{sec:CO maps}. The maps are truncated at 10 $\mu {\rm K}$ for better comparison. We can see that our UNet can effectively recover the CO signal map no matter how many modes are subtracted in the PCA procedure.

For further comparison and assessment, we calculate and show the CO power spectra of PCA preprocessed, UNet generated and signal or target maps in Figure~\ref{fig:power comparison fiducial}. The relative deviation $(P - P_{\rm tar})/P_{\rm tar}$ are also shown in the lower panels. The errors of the power spectra are calculated by root-mean-square (rms) of all the power spectra. We notice that the PCA-1, i.e. the case of removing the first largest mode, cannot effectively remove the foregrounds, resulting in large deviations compared to CO signal power spectrum. Then the first two or three largest modes need to be removed (i.e. PCA-2 and PCA-3) for correctly removing the foregrounds. However, we find that the signal loss effect still can be clearly recognized at large and small scales induced by PCA and beam effect as expected. As we mentioned, although the PCA blind subtraction can remove most continuum foregrounds from CO intensity maps, it also can corrupt the information at large scale, making deviations in power spectra compared to the signal. Usually, the more modes PCA subtracts, the more signal loses at large scales, and the beam effect further complicates this situation that information at both large and small scales will be corrupted because of the correlation of beam and frequencies. On the other hand, in Figure~\ref{fig:power comparison fiducial}, we can find that our UNet can excellently recover the power spectra regardless the number of modes subtracted by PCA, even in the non-linear regime at $k\gtrsim1\, {\rm Mpc^{-1}}h$. The deviations at large and small scale induced by PCA and beam effect can be corrected, and all derived data points are consistent with the CO signal power spectrum within 1$\sigma$ confidence level.

For the low CO signal model with CO signal approximately 100 times lower (see Figure~\ref{fig:IR_CO_relation}), we find that more modes of PCA are required to clean the foregrounds. Similar to Figure~\ref{fig:map comparison fiducial} and \ref{fig:power comparison fiducial}, We show the relevant maps and power spectra of the low CO signal model at 57.6 GHz in Figure~\ref{fig:map comparison low} and Figure~\ref{fig:power comparison low}, respectively. The maps are truncated at 0.1 $\mu$K for better comparison. As can be seen, the PCA-1 and PCA-2 cannot effectively remove the foregrounds in this case, and the web structures from the CO signal are still deeply buried in foreground residuals. This is also indicated in Figure~\ref{fig:power comparison low}, that there is large discrepancy between the power spectra of PCA-1 or PCA-2 (blue data points) and the CO signal (orange curves). Our UNet manages to extract the signal from the PCA-1 map, and some deviation still exists in the derived power spectrum (green data points) compared to the CO signal power spectrum. For the PCA-2 maps, the UNet can excellently recover the signals both in the derived maps and power spectrum, which is consistent with the CO signal power spectrum in 1$\sigma$. In the PCA-3 case, although the result after the mode subtraction procedure is much better than the PCA-1 and PCA-2 cases, the signal loss effect is still strong at large scales. Here our UNet can perfectly reconstruct the CO signal in the derived map, and recover the signal power spectra even better than the result of the PCA-2 case.

\begin{figure}
	\includegraphics[width=\columnwidth]{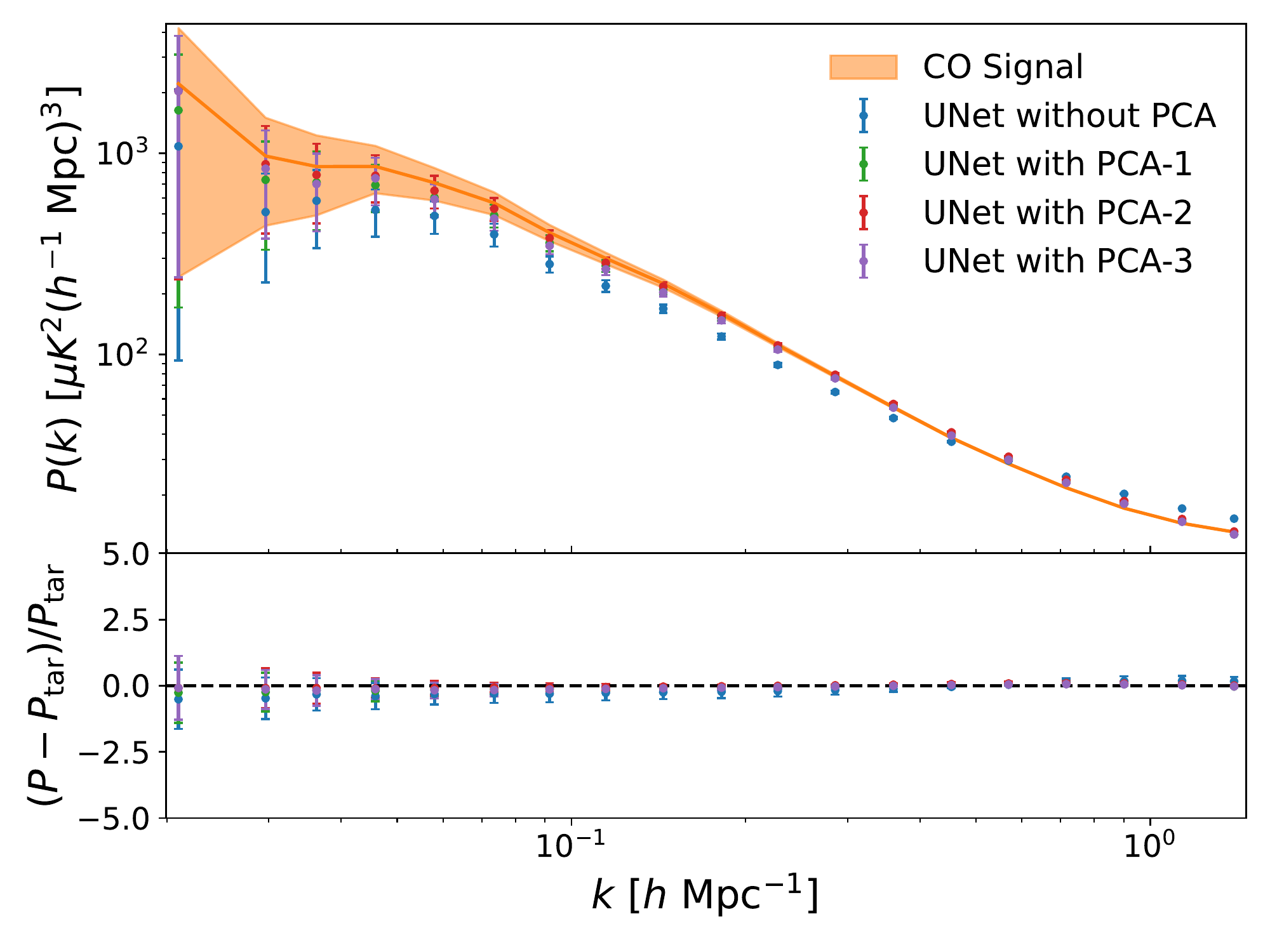}
	\caption{Comparison between the power spectra of the maps generated by UNet with (green, red and purple data points) and without (blue data points) PCA preprocessing for the case of our fiducial CO model. We notice that the power spectrum reconstructed by the network suffers deviations at all scales without PCA preprocessing, and it is even worse in the low CO signal model. Hence, PCA is a necessary procedure in our neural network analysis.}
	\label{fig:comparison pca and no pca}
\end{figure}

We also investigate the cases without PCA preprocessing. For our fiducial CO model, we find that the UNet can generate quite similar CO maps with more training epochs compared to the result with PCA preprocessing. The derived CO power spectra with and without PCA preprocessing  are shown in Figure~\ref{fig:comparison pca and no pca}. We notice that the power spectrum derived from the network without PCA preprocessing (blue data points) has more than $1-2\sigma$ deviations compared to the CO signal power spectrum (orange curves) at all scales. On the other hand, the results with PCA preprocessing can match the CO signal power spectrum in 1$\sigma$.
For the case of our low CO signal model, the CO signals is completely dominated by foregrounds, and our UNet cannot derive correct CO map and the power spectrum without PCA preprocessing. Therefore, PCA preprocessing is a necessary procedure in our analysis using the neural network.

In addition, we try to use the neural network to extract the CO signal considering significant instrumental noise. Our assumed instrumental parameters are described in Section~\ref{sec:instrumental effects}, and we generate the Gaussian noise maps and add them into the total CO maps. We find that the amplitude of this noise power spectrum is about 2600 $\mu K^2(h^{-1}{\rm Mpc})^3$, which is higher than the signal power spectrum at all scales for our fiducial CO model (see the power spectra of Figure~\ref{fig:example and ps}). In this case, we cannot obtain correct CO signal maps and power spectra using the neural network, since the CO signal is still totally dominated by the Gaussian noise after the PCA preprocessing. We find that only if the Gaussian noise can be 1/10 or lower compared to the current level, our network can derive the correct results. This indicates that low instrumental noise is required in deep learning analysis for obtaining correct line signal maps and power spectra, which is also the requirement in traditional signal extraction methods. Since the instrumental noise can be reduced by increasing total observational time, adding more instrumental feeds, or decreasing the system temperature, this problem is fortunately solvable in the intensity mapping surveys. Therefore, deep learning could be an effective method in map foreground removal of CO and other line intensity mapping with low instrumental noise, that also has been proven in previous studies \citep[e.g.][]{Moriwaki2021,Ni2022}.

\section{Summary and Conclusion}
\label{sec:conclusion}
In this work, we investigate the foreground removal of CO intensity mapping at $z\simeq1$ by deep learning methods. The CO intensity maps are simulated based on N-body simulations and the relation of CO luminosity and halo mass $L_{\rm CO}-M_{\rm h}$ from CO and IR observations. We derive the upper, median, and lower relations of the current $L_{\rm CO}-L_{\rm IR}$ measurements, and set the median and lower cases as our fiducial and low CO models, respectively. The CO intensity is approximately 100 times smaller for the low CO model  than the fiducial one.

We also include foreground contamination, instrumental noise, and beam effect in the mock data. The foreground components we consider are thermal dust, spinning dust, synchrotron, free-free emissions and CMB anisotropy at frequencies of CO signals generated by PySM3 python package. The instrumental effects, such as instrumental noise and beam effects of radio telescope are also considered. The instrumental noise is induced by system temperature and it is assumed to be Gaussian-like. The beam model we investigate is Bessel or Jinc beam, which has the largest sidelobes among models studied in literature. CO maps and foregrounds are all convolved with this beam model to simulate the beam effects. We find the traditional method such as PCA will lead to deviations of power spectra by signal loss effect, and considering the beam effect, it becomes even worse. Then we try to use deep learning for accurately recovering the CO signal map with correct line power spectrum.

Our deep learning model is based on ResUNet, which combines the image-to-image generation UNet with ResNet architecture. Since PCA can effectively remove most foregrounds, we preprocess the contaminated CO maps with PCA before feeding our UNet. We find that for our fiducial CO model, removing the first two or more largest modes (i.e. PCA-2 and PCA-3 in our analysis) in PCA can effectively subtract the foregrounds, but resulting in some deviations in power spectra due to signal loss effect. Our UNet can accurately recover the CO signal maps with correct line power spectra, no matter how many modes are subtracted. For the low CO signal model, at least the first three largest modes (i.e. PCA-3) need to be removed in the PCA procedure for effectively subtracting the foregrounds but with large signal loss. However, we find that our UNet can obtain the correct CO signal maps and recover the line power spectra in both PCA-2 and PCA-3 cases. We also investigate the effects of PCA preprocessing and instrumental noise on the results from deep learning. We find that PCA preprocessing for foreground removal is necessary, especially for low signal case, and the instrumental noise also need to be limited in relatively low levels for accurate signal extraction. Besides, Our method also can be easily applied in other line intensity mapping like neutral hydrogen 21cm emission. Therefore, the deep learning method could be an effective way for foreground removal in CO and other intensity mapping, and can obtain accurate line power spectrum for exploring the cosmic large-scale structure, star formation and galaxy evolution.

\section*{Acknowledgements}

X.C.Z. and Y.G. acknowledge the support of 2020SKA0110402, MOST-2018YFE0120800, National Key R\&D Program of China No.2022YFF0503404, NSFC-11822305, NSFC-11773031, and NSFC-11633004. X.L.C. acknowledges the support of the National Natural Science Foundation of China through grant No. 11473044, 11973047, and the Chinese Academy of Science grants QYZDJ-SSW-SLH017, XDB 23040100, XDA15020200. This work is also supported by the science research grants from the China Manned Space Project with NO.CMS-CSST-2021-B01 and CMS- CSST-2021-A01. This work was funded by the National Natural Science Foundation of China (NSFC) under No.11080922.

\section*{Data Availability}

The data that support the findings of this study are available from the corresponding author, upon reasonable request.



\bibliographystyle{mnras}
\bibliography{example} 

\begin{thebibliography}{}
\makeatletter
\relax
\def\mn@urlcharsother{\let\do\@makeother \do\$\do\&\do\#\do\^\do\_\do\%\do\~}
\def\mn@doi{\begingroup\mn@urlcharsother \@ifnextchar [ {\mn@doi@}
  {\mn@doi@[]}}
\def\mn@doi@[#1]#2{\def\@tempa{#1}\ifx\@tempa\@empty \href
  {http://dx.doi.org/#2} {doi:#2}\else \href {http://dx.doi.org/#2} {#1}\fi
  \endgroup}
\def\mn@eprint#1#2{\mn@eprint@#1:#2::\@nil}
\def\mn@eprint@arXiv#1{\href {http://arxiv.org/abs/#1} {{\tt arXiv:#1}}}
\def\mn@eprint@dblp#1{\href {http://dblp.uni-trier.de/rec/bibtex/#1.xml}
  {dblp:#1}}
\def\mn@eprint@#1:#2:#3:#4\@nil{\def\@tempa {#1}\def\@tempb {#2}\def\@tempc
  {#3}\ifx \@tempc \@empty \let \@tempc \@tempb \let \@tempb \@tempa \fi \ifx
  \@tempb \@empty \def\@tempb {arXiv}\fi \@ifundefined
  {mn@eprint@\@tempb}{\@tempb:\@tempc}{\expandafter \expandafter \csname
  mn@eprint@\@tempb\endcsname \expandafter{\@tempc}}}

\bibitem[\protect\citeauthoryear{Abdi \& Williams}{Abdi \&
  Williams}{2010}]{Abdi2010}
Abdi H.,  Williams L.~J.,  2010, Wiley interdisciplinary reviews: computational
  statistics, 2, 433

\bibitem[\protect\citeauthoryear{Alonso, Bull, Ferreira  \& Santos}{Alonso
  et~al.}{2014}]{Alonso2014}
Alonso D.,  Bull P.,  Ferreira P.~G.,   Santos M.~G.,  2014, \mn@doi [Monthly
  Notices of the Royal Astronomical Society] {10.1093/mnras/stu2474}, 447, 400

\bibitem[\protect\citeauthoryear{{Aragon-Calvo}}{{Aragon-Calvo}}{2019}]{Aragon2019}
{Aragon-Calvo} M.~A.,  2019, \mn@doi [\mnras] {10.1093/mnras/stz393}, \href
  {https://ui.adsabs.harvard.edu/abs/2019MNRAS.484.5771A} {484, 5771}

\bibitem[\protect\citeauthoryear{{Aravena} et~al.,}{{Aravena}
  et~al.}{2012}]{Aravena2012}
{Aravena} M.,  et~al., 2012, \mn@doi [\mnras]
  {10.1111/j.1365-2966.2012.21697.x}, \href
  {https://ui.adsabs.harvard.edu/abs/2012MNRAS.426..258A} {426, 258}

\bibitem[\protect\citeauthoryear{Asad et~al.,}{Asad et~al.}{2021}]{Asad2021}
Asad K.,  et~al., 2021, Monthly Notices of the Royal Astronomical Society, 502,
  2970

\bibitem[\protect\citeauthoryear{{Bernal} \& {Kovetz}}{{Bernal} \&
  {Kovetz}}{2022}]{Bernal2022}
{Bernal} J.~L.,  {Kovetz} E.~D.,  2022, arXiv e-prints, \href
  {https://ui.adsabs.harvard.edu/abs/2022arXiv220615377B} {p. arXiv:2206.15377}

\bibitem[\protect\citeauthoryear{{Breysse} et~al.,}{{Breysse}
  et~al.}{2022}]{Breysse2022}
{Breysse} P.~C.,  et~al., 2022, \mn@doi [\apj] {10.3847/1538-4357/ac63c9},
  \href {https://ui.adsabs.harvard.edu/abs/2022ApJ...933..188B} {933, 188}

\bibitem[\protect\citeauthoryear{{CCAT-Prime collaboration}
  et~al.,}{{CCAT-Prime collaboration} et~al.}{2021}]{Aravena2021}
{CCAT-Prime collaboration} et~al., 2021, arXiv e-prints, \href
  {https://ui.adsabs.harvard.edu/abs/2021arXiv210710364C} {p. arXiv:2107.10364}

\bibitem[\protect\citeauthoryear{{Carilli}}{{Carilli}}{2011}]{Carilli2011}
{Carilli} C.~L.,  2011, \mn@doi [\apjl] {10.1088/2041-8205/730/2/L30}, \href
  {https://ui.adsabs.harvard.edu/abs/2011ApJ...730L..30C} {730, L30}

\bibitem[\protect\citeauthoryear{{Carilli} \& {Walter}}{{Carilli} \&
  {Walter}}{2013}]{Carilli2013}
{Carilli} C.~L.,  {Walter} F.,  2013, \mn@doi [\araa]
  {10.1146/annurev-astro-082812-140953}, \href
  {https://ui.adsabs.harvard.edu/abs/2013ARA&A..51..105C} {51, 105}

\bibitem[\protect\citeauthoryear{{Chung} et~al.,}{{Chung}
  et~al.}{2022}]{Chung2022}
{Chung} D.~T.,  et~al., 2022, \mn@doi [\apj] {10.3847/1538-4357/ac63c7}, \href
  {https://ui.adsabs.harvard.edu/abs/2022ApJ...933..186C} {933, 186}

\bibitem[\protect\citeauthoryear{{Cleary} et~al.,}{{Cleary}
  et~al.}{2022}]{Cleary2022}
{Cleary} K.~A.,  et~al., 2022, \mn@doi [\apj] {10.3847/1538-4357/ac63cc}, \href
  {https://ui.adsabs.harvard.edu/abs/2022ApJ...933..182C} {933, 182}

\bibitem[\protect\citeauthoryear{{Condon} \& {Ransom}}{{Condon} \&
  {Ransom}}{2016}]{Condon2016}
{Condon} J.~J.,  {Ransom} S.~M.,  2016, {Essential Radio Astronomy}

\bibitem[\protect\citeauthoryear{{Daddi} et~al.,}{{Daddi}
  et~al.}{2010}]{Daddi2010}
{Daddi} E.,  et~al., 2010, \mn@doi [\apjl] {10.1088/2041-8205/714/1/L118},
  \href {https://ui.adsabs.harvard.edu/abs/2010ApJ...714L.118D} {714, L118}

\bibitem[\protect\citeauthoryear{{Deng}, {Gong}, {Wang}, {Dong}, {Cao}  \&
  {Chen}}{{Deng} et~al.}{2022}]{Deng2022}
{Deng} F.,  {Gong} Y.,  {Wang} Y.,  {Dong} S.,  {Cao} Y.,   {Chen} X.,  2022,
  \mn@doi [\mnras] {10.1093/mnras/stac2185}, \href
  {https://ui.adsabs.harvard.edu/abs/2022MNRAS.515.5894D} {515, 5894}

\bibitem[\protect\citeauthoryear{{Dessauges-Zavadsky}
  et~al.,}{{Dessauges-Zavadsky} et~al.}{2015}]{Dessauges2015}
{Dessauges-Zavadsky} M.,  et~al., 2015, \mn@doi [\aap]
  {10.1051/0004-6361/201424661}, \href
  {https://ui.adsabs.harvard.edu/abs/2015A&A...577A..50D} {577, A50}

\bibitem[\protect\citeauthoryear{{Diakogiannis}, {Waldner}, {Caccetta}  \&
  {Wu}}{{Diakogiannis} et~al.}{2020}]{Diakogiannis2020}
{Diakogiannis} F.~I.,  {Waldner} F.,  {Caccetta} P.,   {Wu} C.,  2020, \mn@doi
  [ISPRS Journal of Photogrammetry and Remote Sensing]
  {10.1016/j.isprsjprs.2020.01.013}, \href
  {https://ui.adsabs.harvard.edu/abs/2020JPRS..162...94D} {162, 94}

\bibitem[\protect\citeauthoryear{{Feder}, {Berger}  \& {Stein}}{{Feder}
  et~al.}{2020}]{Feder2020}
{Feder} R.~M.,  {Berger} P.,   {Stein} G.,  2020, \mn@doi [\prd]
  {10.1103/PhysRevD.102.103504}, \href
  {https://ui.adsabs.harvard.edu/abs/2020PhRvD.102j3504F} {102, 103504}

\bibitem[\protect\citeauthoryear{{Fonseca}, {Silva}, {Santos}  \&
  {Cooray}}{{Fonseca} et~al.}{2017}]{Fonseca2017}
{Fonseca} J.,  {Silva} M.~B.,  {Santos} M.~G.,   {Cooray} A.,  2017, \mn@doi
  [\mnras] {10.1093/mnras/stw2470}, \href
  {https://ui.adsabs.harvard.edu/abs/2017MNRAS.464.1948F} {464, 1948}

\bibitem[\protect\citeauthoryear{{Fukushima} \& {Miyake}}{{Fukushima} \&
  {Miyake}}{1982}]{Fukushima1982}
{Fukushima} K.,  {Miyake} S.,  1982, \mn@doi [Pattern Recognition]
  {10.1016/0031-3203(82)90024-3}, \href
  {https://ui.adsabs.harvard.edu/abs/1982PatRe..15..455F} {15, 455}

\bibitem[\protect\citeauthoryear{{Gong}, {Cooray}, {Silva}, {Santos}  \&
  {Lubin}}{{Gong} et~al.}{2011}]{Gong2011}
{Gong} Y.,  {Cooray} A.,  {Silva} M.~B.,  {Santos} M.~G.,   {Lubin} P.,  2011,
  \mn@doi [\apjl] {10.1088/2041-8205/728/2/L46}, \href
  {https://ui.adsabs.harvard.edu/abs/2011ApJ...728L..46G} {728, L46}

\bibitem[\protect\citeauthoryear{{Gong}, {Cooray}, {Silva}, {Santos}, {Bock},
  {Bradford}  \& {Zemcov}}{{Gong} et~al.}{2012}]{Gong2012}
{Gong} Y.,  {Cooray} A.,  {Silva} M.,  {Santos} M.~G.,  {Bock} J.,  {Bradford}
  C.~M.,   {Zemcov} M.,  2012, \mn@doi [\apj] {10.1088/0004-637X/745/1/49},
  \href {https://ui.adsabs.harvard.edu/abs/2012ApJ...745...49G} {745, 49}

\bibitem[\protect\citeauthoryear{{Gong}, {Cooray}  \& {Santos}}{{Gong}
  et~al.}{2013}]{Gong2013}
{Gong} Y.,  {Cooray} A.,   {Santos} M.~G.,  2013, \mn@doi [\apj]
  {10.1088/0004-637X/768/2/130}, \href
  {https://ui.adsabs.harvard.edu/abs/2013ApJ...768..130G} {768, 130}

\bibitem[\protect\citeauthoryear{{Gong}, {Silva}, {Cooray}  \& {Santos}}{{Gong}
  et~al.}{2014}]{Gong2014}
{Gong} Y.,  {Silva} M.,  {Cooray} A.,   {Santos} M.~G.,  2014, \mn@doi [\apj]
  {10.1088/0004-637X/785/1/72}, \href
  {https://ui.adsabs.harvard.edu/abs/2014ApJ...785...72G} {785, 72}

\bibitem[\protect\citeauthoryear{Gong, Cooray, Silva, Zemcov, Feng, Santos,
  Dore  \& Chen}{Gong et~al.}{2017}]{Gong2017}
Gong Y.,  Cooray A.,  Silva M.~B.,  Zemcov M.,  Feng C.,  Santos M.~G.,  Dore
  O.,   Chen X.,  2017, \mn@doi [The Astrophysical Journal]
  {10.3847/1538-4357/835/2/273}, 835, 273

\bibitem[\protect\citeauthoryear{Gong, Chen  \& Cooray}{Gong
  et~al.}{2020}]{Gong2020}
Gong Y.,  Chen X.,   Cooray A.,  2020, \mn@doi [The Astrophysical Journal]
  {10.3847/1538-4357/ab87a0}, 894, 152

\bibitem[\protect\citeauthoryear{{Goodfellow}, {Pouget-Abadie}, {Mirza}, {Xu},
  {Warde-Farley}, {Ozair}, {Courville}  \& {Bengio}}{{Goodfellow}
  et~al.}{2014}]{Goodfellow2014}
{Goodfellow} I.~J.,  {Pouget-Abadie} J.,  {Mirza} M.,  {Xu} B.,  {Warde-Farley}
  D.,  {Ozair} S.,  {Courville} A.,   {Bengio} Y.,  2014, arXiv e-prints, \href
  {https://ui.adsabs.harvard.edu/abs/2014arXiv1406.2661G} {p. arXiv:1406.2661}

\bibitem[\protect\citeauthoryear{{Greve} et~al.,}{{Greve}
  et~al.}{2014}]{Greve2014}
{Greve} T.~R.,  et~al., 2014, \mn@doi [\apj] {10.1088/0004-637X/794/2/142},
  \href {https://ui.adsabs.harvard.edu/abs/2014ApJ...794..142G} {794, 142}

\bibitem[\protect\citeauthoryear{Gruppioni et~al.,}{Gruppioni
  et~al.}{2013}]{Gruppioni2013}
Gruppioni C.,  et~al., 2013, \mn@doi [Monthly Notices of the Royal Astronomical
  Society] {10.1093/mnras/stt308}, 432, 23

\bibitem[\protect\citeauthoryear{Harper, Dickinson, Battye, Roychowdhury,
  Browne, Ma, Olivari  \& Chen}{Harper et~al.}{2018}]{Harper2018}
Harper S.~E.,  Dickinson C.,  Battye R.~A.,  Roychowdhury S.,  Browne I. W.~A.,
   Ma Y.-Z.,  Olivari L.~C.,   Chen T.,  2018, \mn@doi [Monthly Notices of the
  Royal Astronomical Society] {10.1093/mnras/sty1238}, 478, 2416

\bibitem[\protect\citeauthoryear{{He}, {Zhang}, {Ren}  \& {Sun}}{{He}
  et~al.}{2015}]{He2015}
{He} K.,  {Zhang} X.,  {Ren} S.,   {Sun} J.,  2015, arXiv e-prints, \href
  {https://ui.adsabs.harvard.edu/abs/2015arXiv151203385H} {p. arXiv:1512.03385}

\bibitem[\protect\citeauthoryear{Hyvärinen \& Oja}{Hyvärinen \&
  Oja}{2000}]{Hyvarinen2000}
Hyvärinen A.,  Oja E.,  2000, \mn@doi [Neural Networks]
  {https://doi.org/10.1016/S0893-6080(00)00026-5}, 13, 411

\bibitem[\protect\citeauthoryear{{Ihle} et~al.,}{{Ihle}
  et~al.}{2022}]{Ihle2022}
{Ihle} H.~T.,  et~al., 2022, \mn@doi [\apj] {10.3847/1538-4357/ac63c5}, \href
  {https://ui.adsabs.harvard.edu/abs/2022ApJ...933..185I} {933, 185}

\bibitem[\protect\citeauthoryear{{Ioffe} \& {Szegedy}}{{Ioffe} \&
  {Szegedy}}{2015}]{Ioffe2015}
{Ioffe} S.,  {Szegedy} C.,  2015, arXiv e-prints, \href
  {https://ui.adsabs.harvard.edu/abs/2015arXiv150203167I} {p. arXiv:1502.03167}

\bibitem[\protect\citeauthoryear{{Isola}, {Zhu}, {Zhou}  \& {Efros}}{{Isola}
  et~al.}{2016}]{Isola2016}
{Isola} P.,  {Zhu} J.-Y.,  {Zhou} T.,   {Efros} A.~A.,  2016, arXiv e-prints,
  \href {https://ui.adsabs.harvard.edu/abs/2016arXiv161107004I} {p.
  arXiv:1611.07004}

\bibitem[\protect\citeauthoryear{{Karkare} et~al.,}{{Karkare}
  et~al.}{2022}]{Karkare2022}
{Karkare} K.~S.,  et~al., 2022, \mn@doi [Journal of Low Temperature Physics]
  {10.1007/s10909-022-02702-2}, \href
  {https://ui.adsabs.harvard.edu/abs/2022JLTP..tmp...61K} {}

\bibitem[\protect\citeauthoryear{{Kasmanoff}, {Villaescusa-Navarro}, {Tinker}
  \& {Ho}}{{Kasmanoff} et~al.}{2020}]{Kasmanoff2020}
{Kasmanoff} N.,  {Villaescusa-Navarro} F.,  {Tinker} J.,   {Ho} S.,  2020,
  arXiv e-prints, \href {https://ui.adsabs.harvard.edu/abs/2020arXiv201200186K}
  {p. arXiv:2012.00186}

\bibitem[\protect\citeauthoryear{{Keating} et~al.,}{{Keating}
  et~al.}{2015}]{Keating2015}
{Keating} G.~K.,  et~al., 2015, \mn@doi [\apj] {10.1088/0004-637X/814/2/140},
  \href {https://ui.adsabs.harvard.edu/abs/2015ApJ...814..140K} {814, 140}

\bibitem[\protect\citeauthoryear{{Keating}, {Marrone}, {Bower}, {Leitch},
  {Carlstrom}  \& {DeBoer}}{{Keating} et~al.}{2016}]{Keating2016}
{Keating} G.~K.,  {Marrone} D.~P.,  {Bower} G.~C.,  {Leitch} E.,  {Carlstrom}
  J.~E.,   {DeBoer} D.~R.,  2016, \mn@doi [\apj] {10.3847/0004-637X/830/1/34},
  \href {https://ui.adsabs.harvard.edu/abs/2016ApJ...830...34K} {830, 34}

\bibitem[\protect\citeauthoryear{{Keating}, {Marrone}, {Bower}  \&
  {Keenan}}{{Keating} et~al.}{2020}]{Keating2020}
{Keating} G.~K.,  {Marrone} D.~P.,  {Bower} G.~C.,   {Keenan} R.~P.,  2020,
  \mn@doi [\apj] {10.3847/1538-4357/abb08e}, \href
  {https://ui.adsabs.harvard.edu/abs/2020ApJ...901..141K} {901, 141}

\bibitem[\protect\citeauthoryear{{Kingma} \& {Ba}}{{Kingma} \&
  {Ba}}{2014}]{Kingma2014}
{Kingma} D.~P.,  {Ba} J.,  2014, arXiv e-prints, \href
  {https://ui.adsabs.harvard.edu/abs/2014arXiv1412.6980K} {p. arXiv:1412.6980}

\bibitem[\protect\citeauthoryear{{Klypin}, {Yepes}, {Gottl{\"o}ber}, {Prada}
  \& {He{\ss}}}{{Klypin} et~al.}{2016}]{Klypin2016}
{Klypin} A.,  {Yepes} G.,  {Gottl{\"o}ber} S.,  {Prada} F.,   {He{\ss}} S.,
  2016, \mn@doi [\mnras] {10.1093/mnras/stw248}, \href
  {https://ui.adsabs.harvard.edu/abs/2016MNRAS.457.4340K} {457, 4340}

\bibitem[\protect\citeauthoryear{{Kodi Ramanah}, {Charnock},
  {Villaescusa-Navarro}  \& {Wandelt}}{{Kodi Ramanah} et~al.}{2020}]{Kodi2020}
{Kodi Ramanah} D.,  {Charnock} T.,  {Villaescusa-Navarro} F.,   {Wandelt}
  B.~D.,  2020, \mn@doi [\mnras] {10.1093/mnras/staa1428}, \href
  {https://ui.adsabs.harvard.edu/abs/2020MNRAS.495.4227K} {495, 4227}

\bibitem[\protect\citeauthoryear{{Kovetz} et~al.,}{{Kovetz}
  et~al.}{2017}]{Kovetz2017}
{Kovetz} E.~D.,  et~al., 2017, arXiv e-prints, \href
  {https://ui.adsabs.harvard.edu/abs/2017arXiv170909066K} {p. arXiv:1709.09066}

\bibitem[\protect\citeauthoryear{{Li}, {Ni}, {Croft}, {Di Matteo}, {Bird}  \&
  {Feng}}{{Li} et~al.}{2021}]{Li2021}
{Li} Y.,  {Ni} Y.,  {Croft} R. A.~C.,  {Di Matteo} T.,  {Bird} S.,   {Feng} Y.,
   2021, \mn@doi [Proceedings of the National Academy of Science]
  {10.1073/pnas.2022038118}, \href
  {https://ui.adsabs.harvard.edu/abs/2021PNAS..11822038L} {118, e2022038118}

\bibitem[\protect\citeauthoryear{{Lidz} \& {Taylor}}{{Lidz} \&
  {Taylor}}{2016}]{Lidz2016}
{Lidz} A.,  {Taylor} J.,  2016, \mn@doi [\apj] {10.3847/0004-637X/825/2/143},
  \href {https://ui.adsabs.harvard.edu/abs/2016ApJ...825..143L} {825, 143}

\bibitem[\protect\citeauthoryear{{Lidz}, {Furlanetto}, {Oh}, {Aguirre},
  {Chang}, {Dor{\'e}}  \& {Pritchard}}{{Lidz} et~al.}{2011}]{Lidz2011}
{Lidz} A.,  {Furlanetto} S.~R.,  {Oh} S.~P.,  {Aguirre} J.,  {Chang} T.-C.,
  {Dor{\'e}} O.,   {Pritchard} J.~R.,  2011, \mn@doi [\apj]
  {10.1088/0004-637X/741/2/70}, \href
  {https://ui.adsabs.harvard.edu/abs/2011ApJ...741...70L} {741, 70}

\bibitem[\protect\citeauthoryear{Maas, Hannun, Ng  et~al.}{Maas
  et~al.}{2013}]{Maas2013}
Maas A.~L.,  Hannun A.~Y.,  Ng A.~Y.,   et~al., 2013, in Proc. icml. p.~3

\bibitem[\protect\citeauthoryear{Magnelli, Elbaz, Chary, Dickinson, Borgne,
  Frayer  \& Willmer}{Magnelli et~al.}{2009}]{Magnelli2009}
Magnelli B.,  Elbaz D.,  Chary R.,  Dickinson M.,  Borgne D.,  Frayer D.,
  Willmer C.,  2009, \mn@doi [A&A] {10.1051/0004-6361:200811443}, 496

\bibitem[\protect\citeauthoryear{{Magnelli} et~al.,}{{Magnelli}
  et~al.}{2014}]{Magnelli2014}
{Magnelli} B.,  et~al., 2014, \mn@doi [\aap] {10.1051/0004-6361/201322217},
  \href {https://ui.adsabs.harvard.edu/abs/2014A&A...561A..86M} {561, A86}

\bibitem[\protect\citeauthoryear{{Makinen}, {Lancaster}, {Villaescusa-Navarro},
  {Melchior}, {Ho}, {Perreault-Levasseur}  \& {Spergel}}{{Makinen}
  et~al.}{2021}]{Makinen2021}
{Makinen} T.~L.,  {Lancaster} L.,  {Villaescusa-Navarro} F.,  {Melchior} P.,
  {Ho} S.,  {Perreault-Levasseur} L.,   {Spergel} D.~N.,  2021, \mn@doi [\jcap]
  {10.1088/1475-7516/2021/04/081}, \href
  {https://ui.adsabs.harvard.edu/abs/2021JCAP...04..081M} {2021, 081}

\bibitem[\protect\citeauthoryear{{Matshawule}, {Spinelli}, {Santos}  \&
  {Ngobese}}{{Matshawule} et~al.}{2021}]{Matshawule2021}
{Matshawule} S.~D.,  {Spinelli} M.,  {Santos} M.~G.,   {Ngobese} S.,  2021,
  \mn@doi [\mnras] {10.1093/mnras/stab1688}, \href
  {https://ui.adsabs.harvard.edu/abs/2021MNRAS.506.5075M} {506, 5075}

\bibitem[\protect\citeauthoryear{{Mirza} \& {Osindero}}{{Mirza} \&
  {Osindero}}{2014}]{Mirza2014}
{Mirza} M.,  {Osindero} S.,  2014, arXiv e-prints, \href
  {https://ui.adsabs.harvard.edu/abs/2014arXiv1411.1784M} {p. arXiv:1411.1784}

\bibitem[\protect\citeauthoryear{{Moriwaki} \& {Yoshida}}{{Moriwaki} \&
  {Yoshida}}{2021}]{Moriwaki20212}
{Moriwaki} K.,  {Yoshida} N.,  2021, \mn@doi [\apjl]
  {10.3847/2041-8213/ac3cc0}, \href
  {https://ui.adsabs.harvard.edu/abs/2021ApJ...923L...7M} {923, L7}

\bibitem[\protect\citeauthoryear{{Moriwaki}, {Shirasaki}  \&
  {Yoshida}}{{Moriwaki} et~al.}{2021}]{Moriwaki2021}
{Moriwaki} K.,  {Shirasaki} M.,   {Yoshida} N.,  2021, \mn@doi [\apjl]
  {10.3847/2041-8213/abd17f}, \href
  {https://ui.adsabs.harvard.edu/abs/2021ApJ...906L...1M} {906, L1}

\bibitem[\protect\citeauthoryear{{Ni}, {Li}, {Gao}  \& {Zhang}}{{Ni}
  et~al.}{2022}]{Ni2022}
{Ni} S.,  {Li} Y.,  {Gao} L.-Y.,   {Zhang} X.,  2022, \mn@doi [\apj]
  {10.3847/1538-4357/ac7a34}, \href
  {https://ui.adsabs.harvard.edu/abs/2022ApJ...934...83N} {934, 83}

\bibitem[\protect\citeauthoryear{Odena, Dumoulin  \& Olah}{Odena
  et~al.}{2016}]{Odena2016}
Odena A.,  Dumoulin V.,   Olah C.,  2016, \mn@doi [Distill]
  {10.23915/distill.00003}

\bibitem[\protect\citeauthoryear{{Padmanabhan}}{{Padmanabhan}}{2018}]{Padmanabhan2018}
{Padmanabhan} H.,  2018, \mn@doi [\mnras] {10.1093/mnras/stx3250}, \href
  {https://ui.adsabs.harvard.edu/abs/2018MNRAS.475.1477P} {475, 1477}

\bibitem[\protect\citeauthoryear{{Perraudin}, {Srivastava}, {Lucchi},
  {Kacprzak}, {Hofmann}  \& {R{\'e}fr{\'e}gier}}{{Perraudin}
  et~al.}{2019}]{Perraudin2019}
{Perraudin} N.,  {Srivastava} A.,  {Lucchi} A.,  {Kacprzak} T.,  {Hofmann} T.,
   {R{\'e}fr{\'e}gier} A.,  2019, \mn@doi [Computational Astrophysics and
  Cosmology] {10.1186/s40668-019-0032-1}, \href
  {https://ui.adsabs.harvard.edu/abs/2019ComAC...6....5P} {6, 5}

\bibitem[\protect\citeauthoryear{{Planck Collaboration} et~al.,}{{Planck
  Collaboration} et~al.}{2014}]{Ade2014}
{Planck Collaboration} et~al., 2014, \mn@doi [\aap]
  {10.1051/0004-6361/201321591}, \href
  {https://ui.adsabs.harvard.edu/abs/2014A&A...571A..16P} {571, A16}

\bibitem[\protect\citeauthoryear{{Planck Collaboration} et~al.,}{{Planck
  Collaboration} et~al.}{2016}]{Adam2016}
{Planck Collaboration} et~al., 2016, \mn@doi [\aap]
  {10.1051/0004-6361/201525967}, \href
  {https://ui.adsabs.harvard.edu/abs/2016A&A...594A..10P} {594, A10}

\bibitem[\protect\citeauthoryear{{Pullen}, {Dor{\'e}}  \& {Bock}}{{Pullen}
  et~al.}{2014}]{Pullen2014}
{Pullen} A.~R.,  {Dor{\'e}} O.,   {Bock} J.,  2014, \mn@doi [\apj]
  {10.1088/0004-637X/786/2/111}, \href
  {https://ui.adsabs.harvard.edu/abs/2014ApJ...786..111P} {786, 111}

\bibitem[\protect\citeauthoryear{{Rodr{\'\i}guez}, {Kacprzak}, {Lucchi},
  {Amara}, {Sgier}, {Fluri}, {Hofmann}  \&
  {R{\'e}fr{\'e}gier}}{{Rodr{\'\i}guez} et~al.}{2018}]{Rodriguez2018}
{Rodr{\'\i}guez} A.~C.,  {Kacprzak} T.,  {Lucchi} A.,  {Amara} A.,  {Sgier} R.,
   {Fluri} J.,  {Hofmann} T.,   {R{\'e}fr{\'e}gier} A.,  2018, \mn@doi
  [Computational Astrophysics and Cosmology] {10.1186/s40668-018-0026-4}, \href
  {https://ui.adsabs.harvard.edu/abs/2018ComAC...5....4R} {5, 4}

\bibitem[\protect\citeauthoryear{{Ronneberger}, {Fischer}  \&
  {Brox}}{{Ronneberger} et~al.}{2015}]{Ronneberger2015}
{Ronneberger} O.,  {Fischer} P.,   {Brox} T.,  2015, arXiv e-prints, \href
  {https://ui.adsabs.harvard.edu/abs/2015arXiv150504597R} {p. arXiv:1505.04597}

\bibitem[\protect\citeauthoryear{Rumelhart, Hinton  \& Williams}{Rumelhart
  et~al.}{1985}]{Rumelhart1985}
Rumelhart D.~E.,  Hinton G.~E.,   Williams R.~J.,  1985, Technical report,
  Learning internal representations by error propagation.
California Univ San Diego La Jolla Inst for Cognitive Science

\bibitem[\protect\citeauthoryear{{Sheth} \& {Tormen}}{{Sheth} \&
  {Tormen}}{1999}]{Sheth1999}
{Sheth} R.~K.,  {Tormen} G.,  1999, \mn@doi [\mnras]
  {10.1046/j.1365-8711.1999.02692.x}, \href
  {https://ui.adsabs.harvard.edu/abs/1999MNRAS.308..119S} {308, 119}

\bibitem[\protect\citeauthoryear{{Silva}, {Santos}, {Gong}, {Cooray}  \&
  {Bock}}{{Silva} et~al.}{2013}]{Silva2013}
{Silva} M.~B.,  {Santos} M.~G.,  {Gong} Y.,  {Cooray} A.,   {Bock} J.,  2013,
  \mn@doi [\apj] {10.1088/0004-637X/763/2/132}, \href
  {https://ui.adsabs.harvard.edu/abs/2013ApJ...763..132S} {763, 132}

\bibitem[\protect\citeauthoryear{{Silva}, {Santos}, {Cooray}  \&
  {Gong}}{{Silva} et~al.}{2015}]{Silva2015}
{Silva} M.,  {Santos} M.~G.,  {Cooray} A.,   {Gong} Y.,  2015, \mn@doi [\apj]
  {10.1088/0004-637X/806/2/209}, \href
  {https://ui.adsabs.harvard.edu/abs/2015ApJ...806..209S} {806, 209}

\bibitem[\protect\citeauthoryear{{Springel} et~al.,}{{Springel}
  et~al.}{2005}]{Springel2005}
{Springel} V.,  et~al., 2005, \mn@doi [\nat] {10.1038/nature03597}, \href
  {https://ui.adsabs.harvard.edu/abs/2005Natur.435..629S} {435, 629}

\bibitem[\protect\citeauthoryear{{Sun} et~al.,}{{Sun} et~al.}{2022}]{Sun2022}
{Sun} S.,  et~al., 2022, \mn@doi [Research in Astronomy and Astrophysics]
  {10.1088/1674-4527/ac684d}, \href
  {https://ui.adsabs.harvard.edu/abs/2022RAA....22f5020S} {22, 065020}

\bibitem[\protect\citeauthoryear{{Thorne}, {Dunkley}, {Alonso}  \&
  {N{\ae}ss}}{{Thorne} et~al.}{2017}]{Thorne2017}
{Thorne} B.,  {Dunkley} J.,  {Alonso} D.,   {N{\ae}ss} S.,  2017, \mn@doi
  [\mnras] {10.1093/mnras/stx949}, \href
  {https://ui.adsabs.harvard.edu/abs/2017MNRAS.469.2821T} {469, 2821}

\bibitem[\protect\citeauthoryear{{Uzgil}, {Aguirre}, {Bradford}  \&
  {Lidz}}{{Uzgil} et~al.}{2014}]{Uzgil2014}
{Uzgil} B.~D.,  {Aguirre} J.~E.,  {Bradford} C.~M.,   {Lidz} A.,  2014, \mn@doi
  [\apj] {10.1088/0004-637X/793/2/116}, \href
  {https://ui.adsabs.harvard.edu/abs/2014ApJ...793..116U} {793, 116}

\bibitem[\protect\citeauthoryear{Visbal \& Loeb}{Visbal \&
  Loeb}{2010}]{Visbal2010}
Visbal E.,  Loeb A.,  2010, \mn@doi [Journal of Cosmology and Astroparticle
  Physics] {10.1088/1475-7516/2010/11/016}, 2010, 016

\bibitem[\protect\citeauthoryear{{Visbal}, {Trac}  \& {Loeb}}{{Visbal}
  et~al.}{2011}]{Visbal2011}
{Visbal} E.,  {Trac} H.,   {Loeb} A.,  2011, \mn@doi [\jcap]
  {10.1088/1475-7516/2011/08/010}, \href
  {https://ui.adsabs.harvard.edu/abs/2011JCAP...08..010V} {2011, 010}

\bibitem[\protect\citeauthoryear{{Wilson}, {Rohlfs}  \&
  {H{\"u}ttemeister}}{{Wilson} et~al.}{2013}]{Wilson2013}
{Wilson} T.~L.,  {Rohlfs} K.,   {H{\"u}ttemeister} S.,  2013, {Tools of Radio
  Astronomy}, \mn@doi{10.1007/978-3-642-39950-3.
}

\bibitem[\protect\citeauthoryear{Wold, Esbensen  \& Geladi}{Wold
  et~al.}{1987}]{Wold1987}
Wold S.,  Esbensen K.,   Geladi P.,  1987, Chemometrics and intelligent
  laboratory systems, 2, 37

\bibitem[\protect\citeauthoryear{{Yue}, {Ferrara}, {Pallottini}, {Gallerani}
  \& {Vallini}}{{Yue} et~al.}{2015}]{Yue2015}
{Yue} B.,  {Ferrara} A.,  {Pallottini} A.,  {Gallerani} S.,   {Vallini} L.,
  2015, \mn@doi [\mnras] {10.1093/mnras/stv933}, \href
  {https://ui.adsabs.harvard.edu/abs/2015MNRAS.450.3829Y} {450, 3829}

\bibitem[\protect\citeauthoryear{{Zhang}, {Wang}, {Zhang}, {Sun}, {He},
  {Contardo}, {Villaescusa-Navarro}  \& {Ho}}{{Zhang} et~al.}{2019}]{Zhang2019}
{Zhang} X.,  {Wang} Y.,  {Zhang} W.,  {Sun} Y.,  {He} S.,  {Contardo} G.,
  {Villaescusa-Navarro} F.,   {Ho} S.,  2019, arXiv e-prints, \href
  {https://ui.adsabs.harvard.edu/abs/2019arXiv190205965Z} {p. arXiv:1902.05965}

\makeatother
\end{thebibliography}








\bsp	
\label{lastpage}
\end{document}